\documentclass[twocolumn,trackchanges]{aastex7}

\usepackage{CJKutf8}

\usepackage[caption=false]{subfig}
\usepackage{tabularx}
\usepackage{capt-of}
\usepackage{hhline}
\usepackage{amsmath}	
\usepackage{amssymb}	
\usepackage{textcomp}
\usepackage{upgreek}
\usepackage{listings}
\newcommand{\Msun}{M_{\odot}}

\newcommand{\cmark}{\ding{51}} 
\newcommand{\xmark}{\ding{55}} 
\usepackage{pifont} 

\begin{document}

\title{A Novel Formation Channel for Supermassive Black Hole Binaries in the Early Universe via Primordial Black Holes}

\author[orcid=0000-0003-1541-177X,gname=Saiyang,sname=Zhang]{Saiyang Zhang}
\affiliation{Department of Physics, University of Texas at Austin, Austin, TX 78712, USA}
\affiliation{Weinberg Institute for Theoretical Physics, Texas Center for Cosmology and Astroparticle Physics, \\ University of Texas at Austin, Austin, TX 78712, USA}
\email[show]{szhangphys@utexas.edu}

\author[orcid=0000-0002-4966-7450,gname=Boyuan, sname=Liu]{Boyuan Liu} 
\affiliation{Institute of Astronomy, University of Cambridge, Madingley Road, Cambridge, CB3 0HA, UK}
\affiliation{Universit\"at Heidelberg, Zentrum fur Astronomie, Institut f\"ur Theoretische Astrophysik, D-69120 Heidelberg, Germany}
\email[show]{boyuan.liu@uni-heidelberg.de}

\author[orcid=0000-0003-0212-2979,gname=Volker,sname=Bromm]{Volker Bromm}
\affiliation{Weinberg Institute for Theoretical Physics, Texas Center for Cosmology and Astroparticle Physics, \\ University of Texas at Austin, Austin, TX 78712, USA}
\affiliation{Department of Astronomy, University of Texas at Austin, Austin, TX 78712, USA}
\email{fakeemail3@google.com}

\begin{abstract}
We present a novel formation channel for supermassive black hole (SMBH) binaries in the early Universe, driven by primordial black holes (PBHs). Using high-resolution hydrodynamical simulations, we explore the role of massive PBHs ($m_{\rm BH} \sim 10^6~\Msun$) in catalyzing the formation of direct-collapse black holes (DCBHs), providing a natural \textit{in situ} pathway for binary SMBH formation. PBHs enhance local overdensities, accelerate structure formation, and exert thermal feedback on the surrounding medium via accretion. Lyman-Werner (LW) radiation from accreting PBHs suppresses H$_2$ cooling, shifting the dominant gas coolant to atomic hydrogen. When combined with significant baryon–dark matter streaming velocities ($v_{\rm b\chi} \gtrsim 0.8 ~\sigma_{\rm b\chi}$, where $\sigma_{\rm b\chi}$ is the root-mean-square streaming velocity), 
these effects facilitate the formation of dense, gravitationally unstable atomically-cooling gas clouds in the PBH’s wake. These clouds exhibit sustained high inflow rates ($\dot{M}_{\rm infall} \gtrsim 0.1-0.01~\Msun$\,yr$^{-1}$), 
providing ideal conditions for DCBH formation from rapidly growing supermassive stars of $\sim 10^5~\Msun$ 
at redshift $z\sim 20-10$. The resulting systems form SMBH binaries with initial mass ratios $q \sim \mathcal{O}(0.1)$ and separations of $\sim 10$ pc. Such PBH-DCBH binaries provide testable predictions for JWST and ALMA, potentially explaining select high-$z$ sources like Little Red Dots (LRDs), and represent gravitational wave sources for future missions like LISA and TianQin—bridging early-Universe black hole physics, multi-messenger astronomy, and dark matter theory.
\end{abstract}



\section{Introduction}\label{sec:intro} 
 
The unprecedented capabilities of the James Webb Space Telescope (JWST) have revolutionized our understanding of the early Universe, unveiling a substantial population of supermassive black holes (SMBHs) at high redshifts of $z\gtrsim 7$ \citep[e.g.,][]{Goulding2023ApJ...955L..24G, Larson_2023_BH, Bogdan:2023UHZ1, Greene2024, Kovacs2024ApJ...965L..21K, Maiolino2024A&A, Natarajan:2023UHZ1, GHZ9Napolitano2025ApJ...989...75N, Maiolino2025arXiv250522567M}. Among the most intriguing findings from JWST are ``Little Red Dots'' (LRDs), compact sources identified at $4<z<9$, whose nature remains enigmatic~\citep[e.g.,][]{Labbe2023Natur.616..266L, LabbeLRD2025ApJ...978...92L, KokorvLRD2024ApJ...968...38K,  Leung2024:LRDarXiv, Taylor2025ApJ...986..165T, KocevskiLRD2025ApJ...986..126K}. Concurrently, JWST observations also report instances of galaxy mergers at high redshift $z\gtrsim 6$ \citep{UblerAGN2024MNRAS.531..355U, Matsuoka2024ApJ...965L...4M}, highlighting dynamic environments capable of hosting SMBH binaries even in the early Universe. Recently, binary SMBHs have been proposed to explain the unusual spectral characteristics of LRDs \citep{InayoshiBinaryBH2025arXiv}, suggesting a direct observational link between high- and lower-redshift binary black hole populations.

Conventionally, within the $\Lambda$CDM cosmological framework, SMBH binaries are understood to originate from both \textit{in situ} and \textit{ex situ} channels. The \textit{in situ} channel involves the fragmentation of the pristine gas cloud into massive star clusters, and merging stellar remnants will eventually form binary massive BHs~\citep[e.g.,][]{Hirano2018ApJ...855...17H, Latif2020ApJSMBHB,Woods2021}. On the other hand, the \textit{ex situ} channel involves hierarchical galaxy mergers~\citep[e.g.,][]{White1978MNRAS.183..341W, White1991ApJ...379...52W}, as a standard scenario for massive galaxy growth. If each merging galaxy harbors a central SMBH that is massive enough \citep[$\gtrsim 10^8\ M_\odot$,][]{Ma2021}, dynamical friction efficiently removes angular momentum, causing black holes to migrate toward the galaxy center and form gravitationally bound binary systems on parsec-scale separations \citep[e.g.,][]{BegelmanSMBHB1980Nature, ValtaojaSMBHB1989ApJ...343...47V, MilosavljevicFinalpc2003AIPC, MilosavljevicMBHB2003ApJ}. 

Those SMBH binaries are proposed to contribute significantly to the gravitational wave background (GWB) signals~\citep{Hobbs2017NSRev...4..707H,Romano2017LRR....20....2R}, detectable by current Pulsar Timing Arrays (PTAs) at nanohertz frequencies~\citep[e.g.,][]{Sesana2013MNRAS,Nanograv2023ApJ,CPTA2023RAA, PPTA2023ApJ, EPTA2023A&A}. Detecting and characterizing these signals will provide crucial insights into the SMBH formation mechanisms, evolution pathways, and merger rates throughout cosmic time. Furthermore, these SMBH binaries can serve as powerful electromagnetic wave emitters, observable across multiple wavelengths from optical to X-rays~\citep[e.g.,][]{RoedigSMBHB2014ApJ,PopovicSMBBH2012NewAR..56...74P,Westernacher-Schneider2022PhRvD.106j3010W}, enriching our understanding of their environments, accretion processes, and host galaxy properties in the local Universe.

Motivated by both observational and theoretical developments, we propose a novel formation channel for SMBH binaries via the direct-collapse mechanism induced by primordial black holes (PBHs). Direct-collapse black holes (DCBHs) represent astrophysically-seeded black holes, emerging in the early Universe, under peculiar conditions for violent gravitational collapse of metal-poor gas clouds 
\citep{Loeb1994ApJ...432...52L,BrommDCBH2003ApJ...596...34B, Begelman2006:DCBH, Lodato:2006DCBH, Reisswig2013_DCBH, Suazo2019SMBH, Latif2020ApJSMBHB, Chon2020SMS}. On the other hand, PBHs constitute one of the well-motivated dark matter candidates \citep[for a general review, see][]{Carr2020ARNPS..70..355C, Carr2021}, theorized to form shortly after the Big Bang via the collapse of overdense regions~\citep{Zeldovich1967SvA....10..602Z, hawking1971gravitationally,Carr1975ApJ...201....1C, Belotsky2019, Escriva2022}. Previous investigations have explored the impact of PBHs on cosmic thermal history \citep[e.g.,][]{Ricotti2008ApJII,Ali-Haimoud2017PhRvD, Deluca2020JCAP...06..044D, Lu2021, Ziparo2022,Zhang2024MNRAS.528..180Z, Casanueva-Villarreal2025A&A...699A..49C}, structure formation \citep[e.g.,][]{Meszaros1975A&A....38....5M, Afshordi2003ApJ...594L..71A, Kashlinsky2021PhRvL.126a1101K, Cappelluti2022ApJ, Boyuan2023arXiv231204085L, Zhang:2024PBH}, and the seeding of early galaxies and SMBHs \citep[e.g.,][]{Mack2007ApJ, Carr2018MNRAS.478.3756C, Inman2019PhRvD.100h3528I, Kohri2022PhRvD.106d3539K, Boyuan2022MNRAS.514.2376L, Boyuan2022ApJ,Lu2024PhRvD.109l3016L, Huang2024PhRvD.110j3540H,Colazo2024, Ziparo2025JCAP...04..040Z, Zhang2025, Matteri2025arXiv250318850M, DayalPBH2025arXiv250608116D, ProlePBH2025arXiv250611233P}. These studies underscore the rich astrophysical phenomena arising within PBH models, intensively examined through both simulations and analytical frameworks. 

Here, we simulate the evolution of the structure around an isolated massive PBH ($\sim 10^6~\Msun$) and study the critical conditions for potential secondary black hole formation. This particular PBH mass scale is motivated by the change in the equation of state during the $e^+ e^-$ annihilation epoch within cosmic thermal history, when the temperature of the universe is $T\sim 1~\rm MeV$ ~\citep{Carr2021PDU....3100755C}. 
Specifically, we explore how the presence of soft-UV, Lyman-Werner (LW) radiation from BH accretion flows interacts with $\rm H_2$ and $\rm H^-$ within the gas cloud through photo-dissociation~\citep{DB1996ApJ...468..269D,Abel1997NewA....2..181A}, thus suppressing their abundance and consequently reducing the gas cooling efficiency. Therefore, different from~\cite{Zhang2025}, under this LW radiation, we focus mainly on the fate of the pristine gas cloud surrounding the PBH and find the condition for runaway collapse of this cloud into a DCBH\footnote{Throughout the paper, the term ``DCBH'' refers generically to any massive black hole above $10^4\ \rm M_\odot$ that forms from rapid collapse of atomic-cooling clouds, not restricted to the classical monolithic collapse scenario. Alternative pathways, such as super-competitive accretion and stellar collisions can produce such massive black holes in the presence of moderate fragmentation \citep[e.g.,][]{Chon2020SMS,Reinoso2023}.}. 
For the cases where secondary black holes form, we make predictions for the evolution of such binary systems and discuss the possible implications for future observations of their electromagnetic and gravitational wave signals.

In Section~\ref{sec:method}, we describe the numerical recipes for our simulations in detail, including the black hole accretion feedback model and criterion for collapsing gas cloud formation. Following the simulation, we analyze the formation of dense cores and the inflow of gas and discuss the criterion for DCBH formation in Section~\ref{sec:Results}. The potential implications on binary black hole formation and possible observational signatures are discussed in Section~\ref{sec:implications}, followed by conclusions drawn in Section~\ref{sec:conclusion}.

In our simulations, we adopt \textit{Planck18} cosmological parameters throughout \citep{Plank2020A&A...641A...6P}: $\Omega_{\rm m} = 0.3111$, $\Omega_{\rm b}=0.04897$, $h = 0.6776$, $\sigma_8 = 0.8102$, $n_{\rm s} = 0.9665$.

\begin{table*}[ht!]
    \centering
    \caption{Summary of key parameters and main results. $L$ is the size of the box in comoving units. $z_{\rm ini}$ is the initial redshift where the simulation starts, and  $z_{\rm col}$ the redshift when collapsing sink particles begin to aggregate around the central PBH \footnote{Different from \cite{Zhang2025}, limited by computational resources, we terminate the simulations at about $z\gtrsim10$ when the timestep becomes extremely small or when star formation takes place in halos faraway from the PBH.}. $N_{\rm eff}$ is the total number of particles within the simulation box. $\epsilon_r$ is the thermal feedback coupling efficiency. 
    $m_{\rm col}$ denotes the total mass of the collapsing cloud that formed in the PBH-hosting halo by the end of the simulation ($z \sim 10$).  
    STR is a flag indicating whether the relative streaming of PDM and gas particles is included (\cmark) or not (\xmark), and the value represents the amplitude with respect to the root-mean-square streaming velocity $\sigma_{\rm b\chi}$. BH\_LW is another flag to control whether we include (\cmark) the local LW feedback from BH accretion or not (\xmark).  }
    \begin{tabular}{ccccccccc}
    \hline
        Run & $L$\,[ckpc] &$z_{\rm ini}$& $z_{\rm col}$  & $N_{\rm eff}$  & $\epsilon_{\rm r}$&  $m_{\rm col} [\Msun]$   & STR & BH\_LW\\

    \hline
    \texttt{CDM}*\footnote{The simulation runs denoted with * are taken from \cite{Zhang2025}.} & 250 & 1100 & - &  $2\times 256^3$  &  -& -  & \xmark & -  \\
    \texttt{PBH\_fd005}*  & 250 & 1100 & - &  $2\times 256^3$  &  0.005& -  & \xmark & \xmark  \\
        \texttt{PDMonly}*  & 250 & 3400 & - & $ 256^3$  & - & - & -  & - \\
    \hline
        \texttt{PBH\_LW\_fd05} & 250 & 1100 & - &  $2\times 256^3$  &  0.05& -  & \xmark & \cmark  \\
        \texttt{PBH\_LW\_fd005} & 250 & 1100 & - &  $2\times 256^3$   &0.005 & - & \xmark & \cmark \\
        \texttt{PBH\_LW\_fd0005} & 250 & 1100 & - &  $2\times 256^3$   &0.0005 &  - & \xmark & \cmark  \\
\hline
        \texttt{PBH\_LW\_wstr\_fd005} & 250 & 1100 & - &  $2\times 256^3$  &  0.005& -  & 0.4 \cmark & \cmark  \\
        \texttt{PBH\_LW\_str\_fd005} & 250 & 1100 & 17.41 &  $2\times 256^3$  &  0.005& $5.23\times 10^4$  & 0.8 \cmark & \cmark  \\
        \texttt{PBH\_LW\_mstr\_fd005} & 250 & 1100 & 11.71 &  $2\times 256^3$  &  0.005& $5.33\times 10^4$  & 1.2 \cmark & \cmark  \\
        \texttt{PBH\_LW\_sstr\_fd005} & 250 & 1100 & 14.96 &  $2\times 256^3$  &  0.005&  $9.76\times 10^4$ & 1.6 \cmark & \cmark  \\
 \hline 

    \hline
    \end{tabular}
    \label{Table:SimParam}
\end{table*}

\section{Methodology} \label{sec:method}

We explore the evolution of gas dynamics around an isolated PBH at high redshift, using the state-of-the-art simulation package \textsc{gizmo} ~\citep{Hopkins2015MNRAS.450...53H}. In Section~\ref{subsec:ini}, we first describe the simulation code and the initial condition settings. Different from previous work, we focus on the LW photons emitted from BH accretion flows with an improved intensity fitting model~\citep{Takhistov2022, Boyuan2022MNRAS.514.2376L}. We then describe the BH accretion model ~\citep{springel2005cosmological, tremmel2017romulus}, and the implementation of our updated feedback prescription in Section~\ref{subsec:acret}. Last, in Section~\ref{subsec:collapse}, we discuss the criterion for the identification of collapsing gas particles as part of a dense gas core (i.e., the potential formation site of a DCBH). For convenience, Table~\ref{Table:SimParam} summarizes the relevant initial conditions and parameters used in our simulations.

\subsection{Simulation and Initial Conditions}\label{subsec:ini}
We implement our simulations with the \textsc{gizmo} code~\citep{Hopkins2015MNRAS.450...53H}, employing the Lagrangian meshless finite-mass (MFM) solver for hydrodynamics combined with a comprehensive primordial chemistry network \citep[see,][]{Bromm2002ApJ...564...23B,Johnson2006MNRAS.366..247J,LiuBromm2018}, and the parallelized Tree+PM gravity solver for N-body dynamics from \textsc{Gadget-3}~\citep{springel2005cosmological}.

We simulate the initial growth of structure around a PBH with a particle dark matter (PDM)-only pathfinder run denoted as \texttt{PDMonly}, from the beginning of the matter dominated era at $z_{\rm eq} = 3400$ to the recombination epoch ($z = 1100$), using the initial conditions generated with  the \textsc{MUSIC} code \citep{hahn2011multi}, placing an isolated PBH at the center of the box~\footnote{Our PBH initial condition generator \textsc{phantom} ~\citep{zhang_2025_17025634} is publicly available on GitHub: \url{https://github.com/Sylvanzsy/pbh_cosmosim_ics} }. The mass of the PBH is denoted by $m_{\rm BH}$, setting it to be $10^6~\Msun$ throughout. The simulation box has a side length of $L\sim 250~\rm ckpc$, with a total of $N_{\rm PDM} = 256^3$ particles to resolve the PDM component\footnote{As discussed in \citealt{Zhang2025}, our simulations represent an effective PBH mass fraction of less than $6\times 10^{-4}$, set by the
ratio of the PBH mass to the total mass enclosed within the simulation volume. This limit is consistent with the existing constraints and agrees with previous simulation results. The strongest constraint arises from CMB $\mu$-distortions under the assumption of Gaussian primordial fluctuations \citep{Chluba2012ApJ...758...76C,Nakama2018PhRvD..97d3525N,Chluba2021ExA....51.1515C,Hooper2024, Pritchard:2025yda}, with future missions like PIXIE expected to improve limits \citep{PIXIE2025JCAP...04..020K}. However, these constraints can be relaxed in scenarios with non-Gaussianity or alternative inflationary dynamics \citep[e.g.,][]{Kawasaki2019PhRvD.100j3521K, Carr2020ARNPS..70..355C}. We emphasize that our study does not assume a global PBH abundance, but instead focuses on the evolution around an isolated PBH.}. Using the Zel'dovich approximation ~\citep{Zeldovich1970A&A.....5...84Z} and the numerical recipes from previous work ~\citep{Ali-Haimoud2017PhRvD, Inman2019PhRvD.100h3528I, Boyuan2023arXiv231204085L, Zhang:2024PBH}, the displacement and velocity perturbations of PDM particles induced by the PBH are calculated and added in the initial box. Different from~\cite{Zhang:2024PBH}, we have improved the calculation of linear and non-linear perturbation growth by adapting the numerical recipe  from~\cite{Jiao2024}. The growth factor of linear perturbations is rescaled to reproduce the long-term growth of halos seeded by isolated
PBHs from the spherical collapse theory \citep{Mack2007ApJ}.

At $z = 1100$, when baryons and photons start to decouple, we include an additional uniform baryon matter field with the same resolution as PDM in the initial condition set up, resulting in a total number of $N = 2\times 256^3$ particles with a mass of $\sim 100(18)~\Msun$ for PDM (gas) particles. With this setting, we both approximate the initial structure of PDM around the PBH, and let the gas collapse onto this PBH-seeded halo. In running these simulations, the softening length of PDM and gas is set to $\epsilon_{\rm PDM} = \epsilon_{\rm gas} \sim 0.01\, L/N_{\rm PDM}^{1/3} \simeq 0.01~ h^{-1}\rm kpc$, and the initial chemical abundances are assigned to the values predicted for the intergalactic medium (IGM) at $z = 1100$, as summarized in ~\cite{Galli2013ARA&A..51..163G}.

In addition, we also consider the relative streaming motion between gas and PDM particles by assigning a universal velocity offset. In our earlier work \citep{Zhang2025}, we found that streaming can enhance, rather than delay, galaxy formation—contrary to the conclusion generally reached within the standard $\Lambda$CDM framework \citep[e.g.,][]{Schauer2019MNRAS.484.3510S, Schauer2023}. This enhancement arises from a displacement in the center of mass (away from the PBH) caused by the relative motion between baryons and dark matter, leading to star formation in the dense gas structures that form in the wake of the PBH. These wake-driven overdensities promote gravitational collapse and thus play a catalytic role in early star formation. To systematically explore the impact of streaming velocity on black hole formation, we implement a velocity offset parameterized as a multiple of the root-mean-square streaming velocity $\sigma_{\rm b\chi} = 30 ~\rm km\,$s$^{-1}$. Specifically, we consider four cases with velocity offsets of 0.4, 0.8, 1.2, and 1.6 $\sigma_{\rm b\chi}$, corresponding to simulation runs labeled as \texttt{PBH\_LW\_wstr\_fd005}, \texttt{PBH\_LW\_str\_fd005}, \texttt{PBH\_LW\_mstr\_fd005} and \texttt{PBH\_LW\_sstr\_fd005}, respectively.

\subsection{Black Hole Accretion and Feedback}\label{subsec:acret}

\begin{figure}
\centering
    \includegraphics[width=1\columnwidth]{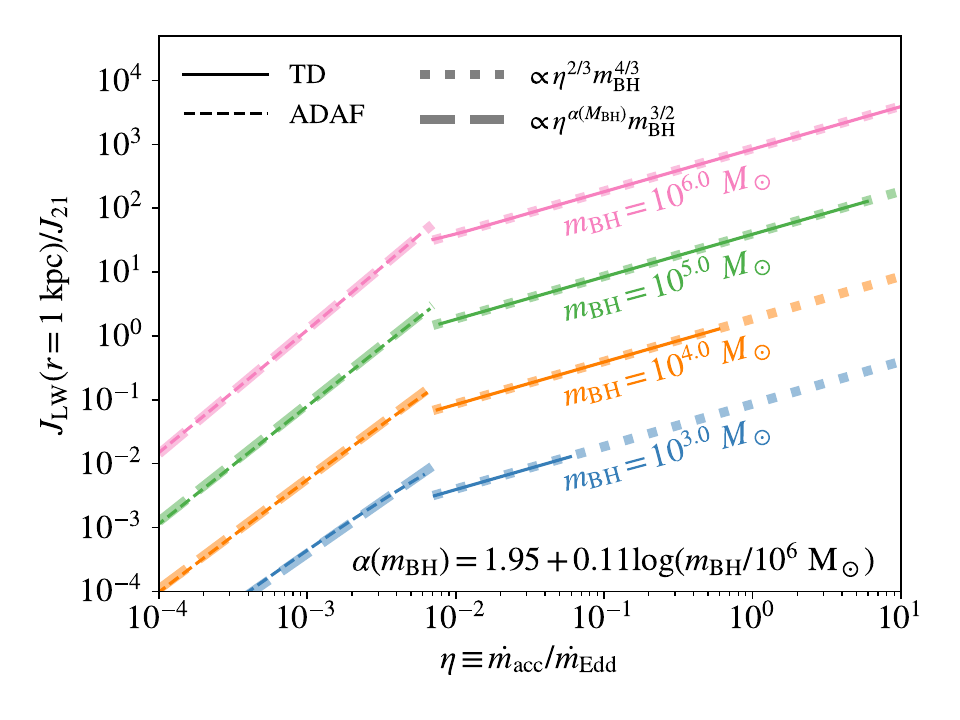}
    \caption{Normalized Lyman–Werner (LW) intensity, $J_{\mathrm{LW}} / J_{21}$, as a function of the Eddington ratio, $\eta\equiv\dot{m}_{\mathrm{acc}} / \dot{m}_{\mathrm{Edd}}$, for black holes of varying masses ($m_{\mathrm{BH}} = 10^{3}, 10^{4}, 10^{5}, 10^{6}\,\Msun$). The $J_{\mathrm{LW}}$ values are computed at a distance of $r = 1~\mathrm{kpc}$ from the black hole. The results based on the semi-analytical spectra models for TD and ADAF accretion profiles in \cite{Takhistov2022} are shown by the thin solid and dashed lines, respectively. Here, the ADAF regime is only expected to occur at low accretion rates ($\eta\lesssim 0.007$), best fitted with  $J_{\mathrm{LW}} \propto \eta^{\alpha(m_{\rm BH})} m_{\mathrm{BH}}^{3/2}$, where $\alpha(m_{\rm BH}) = 1.95+0.11\log\left(m_{\rm BH}/10^6\ \rm M_\odot  \right)$ (Eq.~\ref{eq:jlw}; thick dashed lines).  The TD regime, occurring at higher accretion rates ($\eta > 0.007$), satisfies the scaling relation $J_{\mathrm{LW}} \propto \eta^{2/3} m_{\mathrm{BH}}^{4/3}$ (Eq.~\ref{eq:jlw}; thick dotted lines). 
}
    \label{fig:jlw}
\end{figure}

In the early Universe where the cosmic density field is nearly uniform and isotropic, we use a Bondi-Hoyle formalism to approximate the BH accretion rate, as
\begin{align}
	\dot{m}_{\mathrm{acc}} & = \frac{4\uppi (G m_{\mathrm{BH}})^{2} \rho_{\mathrm{gas}}}{\tilde{v}^{3}} = \frac{4\uppi (G m_{\mathrm{BH}})^{2} \rho_{\mathrm{gas}}}{(c_{s}^{2} + v_{\mathrm{gas}}^{2})^{3/2}} \notag \\
& \simeq 0.0072~\Msun \mathrm{yr}^{-1} \left( \frac{10~\rm km/s}{\tilde{v}} \right)^{3}\notag\\
&\times \left( \frac{n_{\rm H}}{1 \ \rm cm^{-3}} \right) \left(\frac{m_{\mathrm{BH}}}{10^6~\Msun}\right)^{2}, \label{eq:bondi}
\end{align}
where $\rho_{\mathrm{gas}} = \mu m_{\rm H} n_{\rm H}$ is the average density of the gas sampled from the BH accretion kernel, and $\mu = 1.22$ is the average molecular weight. Here, $c_{s}$ and $v_{\mathrm{gas}}$ are the sound speed and velocity dispersion of the surrounding gas, averaged over the gas particles within the accretion kernel. The size of the BH accretion kernel, defined as the region that determines the BH accretion rate, is set to the Bondi radius calculated from the last timestep $r_{\rm Bondi} \sim 2Gm_{\rm BH}/c_s^2$.

From the calculated accretion rate, we update the BH mass at each timestep $\delta t$ by $\delta m_{\mathrm{BH}} = \dot{m}_{\mathrm{acc}} \delta t$. To ensure mass and momentum conservation, we adopt the algorithm from \cite{springel2005cosmological}, in which the BH particle stochastically swallows nearby gas particles to remain consistent with the average growth rate, and a drag force is applied according to the momentum of the swallowed gas. 

During the accretion process, the feedback energy is injected into the surrounding gas particles by a volume-weighted average, following the prescription in \cite{springel2005cosmological} \footnote{Our implementation of PBH feedback assumes isotropic injection of thermal energy, while neglecting mechanical feedback such as disk-driven outflows, collimated jets, or winds. A more detailed treatment incorporating radiative transfer and directional feedback could alter the thermal and dynamical structure of the surrounding gas \citep[see e.g.,][]{Silk2013ApJ...772..112S,Boyuan2023arXiv231204085L}. This as a limitation of the present study and a key direction for future work.}. At the end of each timestep $\delta t$, the total amount of energy injected is $\delta E = \epsilon_r \epsilon_{\rm EM}\dot{m}_{\mathrm{acc}} c^2 \delta t $. Here, $\epsilon_r$ is the thermal coupling coefficient and we take it as a free parameter to study its variational effect at high redshift. $\epsilon_{\rm EM}$ is the radiative efficiency parameter, calculated according to the subgrid model in \cite{Negri2017MNRAS.467.3475N}, as
\begin{align}
\epsilon_{\mathrm{EM}}=\frac{\epsilon_{0}A\eta}{1+A\eta}\ , A=100\mbox{\ ,}\label{epsilonEM}
\end{align}
capturing the transition from the geometrically thick, radiatively inefficient advection-dominated accretion flow (ADAF) regime to a radiatively efficient thin-disk (TD) one. Here $\eta$ is the Eddington ratio defined by $\eta \equiv \dot{m}_{\rm acc}/\dot{m}_{\rm Edd}$, where the Eddington accretion rate $\dot{m}_{\rm Edd}$ is given by  
\begin{align}
    \dot{m}_{\mathrm{Edd}} = 0.047\ \Msun\ \rm yr^{-1}\ \left( \frac{m_{\mathrm{BH}}}{10^{6}\ \Msun} \right) \left( \frac{\epsilon_{0}}{0.057} \right)^{-1}, \label{eq:mdot_edd}
\end{align}
adopting $\epsilon_{0} = 0.057$ as the radiative efficiency in the thin-disk accretion model for non-spinning BHs considering that PBHs are born with very low spins ($\lesssim 0.01$) in the canonical scenario of Gaussian perturbations \citep[e.g.,][]{DeLuca2019,Mirbabayi2020}.

 To estimate the LW radiation intensity, we use fitting formulae in the form of a broken power-law model to capture both the TD and the ADAF regimes, as illustrated in Figure~\ref{fig:jlw}. Here, the ADAF regime is only expected to occur at low accretion rates ($\eta\lesssim 0.007$), transitioning to the TD regime at higher rates.  The specific LW intensities are obtained by integrating the spectra modeled in \cite{Takhistov2022} within the photon energy range $h\nu\sim 11.2$--$13.6$~eV. We can write the intensity (in units of $J_{\rm 21}=10^{-21}\ \rm erg\ s^{-1}\ cm^{-2}\ sr^{-1}\ Hz^{-1}$) at a distance $r$ from an accreting BH as a function of BH mass $m_{\rm BH}$ and Eddington ratio $\eta$ (see Fig.~\ref{fig:jlw}):
\begin{align}
    \frac{J_{\rm LW}}{J_{21}} \simeq 
    \begin{cases}
        9 \times 10^{5} \eta^{1.95+0.11\log\left(m_{\rm BH}/10^6\ \rm M_\odot  \right)} & \\\left( \frac{m_{\rm BH}}{10^6\ \rm M_\odot} \right)^{3/2} \left( \frac{r}{\rm kpc} \right)^{-2} & (\text{ADAF}), \\[8pt]
        8.8 \times 10^{2} \eta^{2/3} &\\\left( \frac{m_{\rm BH}}{10^6\ \rm M_\odot} \right)^{4/3} \left( \frac{r}{\rm kpc} \right)^{-2} & (\text{TD}).
    \end{cases}\label{eq:jlw}
\end{align}

Once $J_{\rm LW}$ is known, we combine it with the local gas shielding factor \citep{Wolcott2011MNRAS.418..838W} to derive the dissociation rates of $\rm H^{\rm -}$ and $\rm H_2$ following \citet[][see their sec.~2.2]{Zhang2025}. 

\subsection{Gravitational Instability}\label{subsec:collapse}
In previous work~\citep{Zhang2025}, we found that under the effect of LW radiation from BH accretion, a dense gas clump with mass $\mathcal{O}(10^5)\Msun$ was identified in the vicinity of the PBH, evolving along the atomic hydrogen cooling track \citep{OhHaiman2002}. Therefore, to further assess whether this gas clump will collapse and may eventually lead to the formation of a DCBH, we here impose an explicit density threshold criterion, $n_{\rm H} \gtrsim 10^6 \ \rm cm^{-3}$, for the formation of collapsing clouds. This value is close to the maximum density resolved within our simulations. It is also chosen to be larger than the density threshold for cloud collapse ($n_{\rm H} \gtrsim 10^4 ~\rm cm^{-3}$) found in previous work, as the ``Zone of no return''~\citep{InayoshiSMS2014MNRAS}. 

To determine whether a gas particle can partake in the collapse or is engulfed by the central BH, we first calculate the local free-fall time of the gas:
\begin{equation}
t_{\rm ff} = \sqrt{\frac{3 \pi}{32 G \rho_{\rm gas} }} \simeq 0.47~{\rm Myr} \left( \frac{10^4 \ \rm cm^{-3}}{n_{\rm H}} \right)^{1/2},
\end{equation}
using the local gas density $\rho_{\rm gas} $. We calculate this timescale once the critical density threshold is reached and compare it to the time, $t_{\rm survive}$, that the particle survives without being accreted by the PBH. 

If this collapsing particle survives the accretion and is not reheated by black hole thermal feedback, such that it can reach $t_{\rm survive} \gtrsim t_{\rm ff}$, we convert it into a sink particle as part of the collapsing gas cloud. However, once the gas particle fails to meet the density collapse criterion while still registering $t_{\rm survive} \lesssim t_{\rm ff}$, we reset the timer to $t_{\rm survive} = 0$ in the following time step, and will not start the timer unless it again satisfies the density threshold criterion. 
We have also established by numerical experimentation that, during the initial collapse at $z\gtrsim 200$, gas densities could reach as high as $n_{\rm H} \gtrsim 10^6 \ \rm cm^{-3}$ in the vicinity of the PBH, thus formally satisfying the collapse criterion mentioned above\footnote{The initial collapse is induced by the PBH seeding effect, where gas is attracted by the potential well forming overdense regions~\citep[for a similar effect without PBHs, see e.g.,][]{Hirano2015ApJ...814...18H, Ito2024PASJ...76..850I,Qin2025arXiv250613858Q, Cyr2025arXiv250717833C}. During this process, the imbalance between the thermal pressure from BH feedback and the gravitational pressure from the gas cloud results in gas collapsing onto the PBH and formation of dense regions near the PBH, greatly boosting the accretion efficiency \citep[see fig.~2 in][]{Zhang2025}.}. However, subsequent shock waves generated during the accretion will rapidly terminate this efficient accretion phase, smoothing out any dense structures around the PBH that have not been accreted. Therefore, to avoid numerical artifacts, we prohibit the formation of collapsing particles as an additional constraint at $z\gtrsim 200$.

To facilitate our analysis, we introduce $N_{\rm col}$ as the cumulative number of sink particles and start to output snapshots when $N_{\rm col} = 1$. Whenever $N_{\rm col}$ doubles, a simulation snapshot is generated to record the detailed time evolution of this dense atomically-cooling gas cloud. Note that the sink particles do not represent individual stars but are only meant to estimate the mass of collapsing gas that would form stars and DCBHs at the limit of our resolution. The detailed star and DCBH formation processes unresolved here are deferred to further work.

\section{Results and Discussion} \label{sec:Results}
In exploring our key results, we first identify the formation of a dense gas core within our simulation box in Section~\ref{subsec:core}. By tracking the inflow of gas towards this dense core, we identify the criterion for the potential gas collapse and subsequent DCBH formation in Section~\ref{subsec:inflowDCBH}.

\subsection{Dense Core Formation}\label{subsec:core}

\begin{figure*}[ht!]
\centering
    \includegraphics[width= \linewidth]{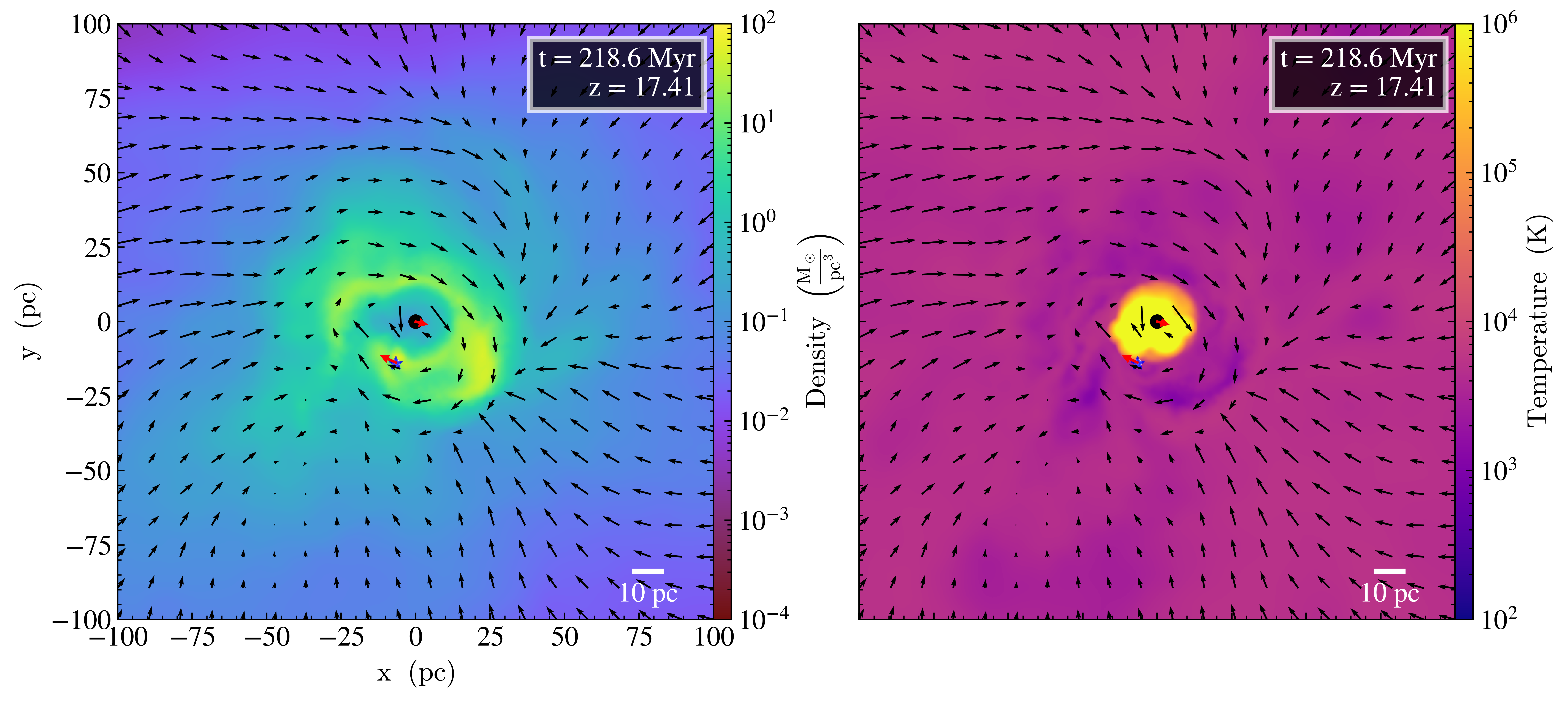}
    \caption{Onset of gaseous cloud collapse. We show projections of gas density (\textbf{left panel}) and temperature (\textbf{right panel}) for the gas surrounding the central PBH taken from the \texttt{PBH\_LW\_str\_fd005} simulation, within a physical 200\,pc scale. The snapshot is taken at $z \simeq 17.4$, corresponding to the moment where the first collapsing sink particle emerges. The simulation assumes a BH thermal feedback efficiency of $\epsilon_r = 0.5\%$, a relative streaming velocity of $v_{\rm b\chi} = 0.8~\sigma_{\rm b\chi}$, and includes the LW radiation generated during PBH accretion. The black dot marks the position of the PBH, while the blue star indicates the location of the collapsing gas cloud, representing the site for potential DCBH formation. Red arrows denote the relative motions between the central PBH and bulk velocity of the collapsing gas cloud. Velocity vectors for gas are overlaid on both panels, with arrow sizes scaled by magnitude (different from that for BHs) to illustrate the inflow of gas toward the PBH, as well as feedback-driven outflows in its vicinity.
 }
    \label{fig:SlicePlot}
\end{figure*}

\begin{figure*}[ht!]
\centering
    \includegraphics[width=  \linewidth]{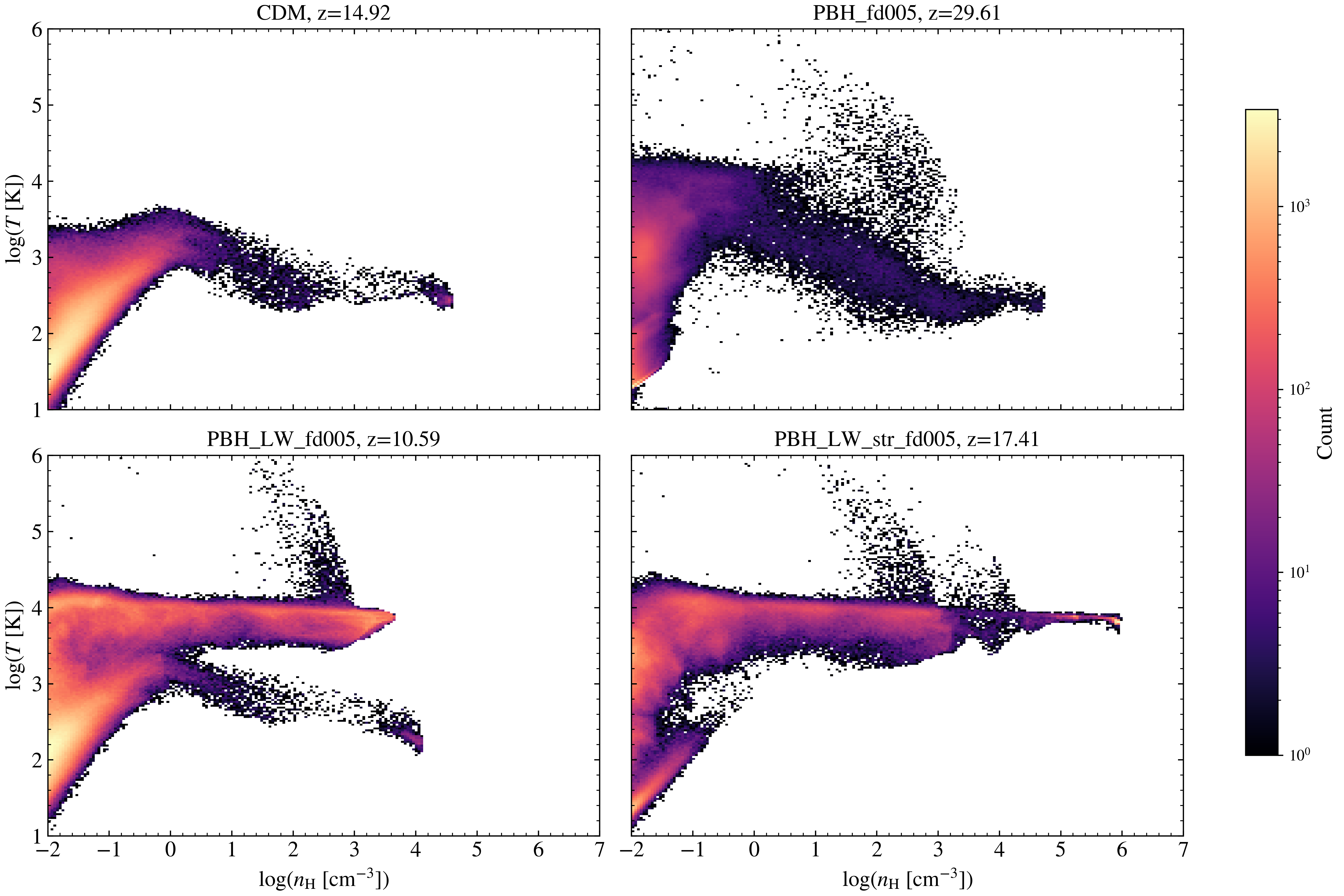}
    \vspace{1ex}
    \caption{Gas properties in the vicinity of the central PBH. We present phase diagrams of temperature ($T$) vs. hydrogen number density ($n_{\mathrm{H}}$) for several simulation runs at the moment where collapsing particles were first identified. The simulation without a PBH (\texttt{CDM}) and without Lyman–Werner (LW) feedback from BH accretion (\texttt{PBH\_fd005}) are included as a reference (taken from \citealt{Zhang2025}) to demonstrate the effects of PBH accretion heating and LW feedback on the surrounding gas. For the runs with PBHs (\texttt{PBH\_fd005},\texttt{PBH\_LW\_fd005} and \texttt{PBH\_LW\_str\_fd005}), the same feedback efficiency of $\epsilon_r = 0.005$ is assumed. The effect of including LW radiation is demonstrated in the lower panels with the \texttt{PBH\_LW\_fd005} (left) and \texttt{PBH\_LW\_str\_fd005} (right) runs, showcasing the change in the thermal evolution of the gas. The additional effect in the presence of baryon–DM streaming is evident in the \texttt{PBH\_LW\_str\_fd005} run, where successful runaway collapse along the near-isothermal atomic cooling track is triggered.
}
    \label{fig:Tn}
\end{figure*}

In the classical DCBH scenario \citep[e.g.,][]{BrommDCBH2003ApJ...596...34B, Begelman2006:DCBH, Lodato:2006DCBH}, the collapse of massive pristine gas clouds (metallicity $Z\lesssim 10^{-4} Z_{\odot}$) occurs under rare conditions that involve strong LW radiation from neighboring galaxies. Under these conditions, the gas gravitationally collapses without significant fragmentation and proto-stellar feedback into an initial protostar at the center \citep{Becerra2018_DCBH,Becerra2018_proto}, rapidly accreting the surrounding gas to form a supermassive star (SMS). This SMS subsequently undergoes instability-triggered collapse into a massive black hole seed of mass $\sim 10^5~\Msun$ at redshift $z\sim 10-15$~\citep[e.g.,][]{Shibata2002_DCBH,Hosokawa2013,Reisswig2013_DCBH,Haemmerle2018,Haemmerle2021,Nagele2022,Herrington2023,Shibata2025}\footnote{Nevertheless, it is now understood that, instead of monolithic collapse, fragmentation is likely unavoidable even in metal-free atomic-cooling clouds, particularly at sub-pc scales due to disk instability \citep{Klessen:2023FirstStars}. It is the competition between fragmentation, protostar growth by accretion, and stellar collisions that determines the seed black hole mass, as discussed in Sec.~\ref{subsec:inflowDCBH}.}

In this study, we introduce a crucial modification to this canonical scenario by incorporating accretion feedback from a central PBH as the LW radiation source, rather than relying on neighboring galaxies. The LW radiation generated by PBH accretion suppresses the abundance of molecular hydrogen (H$_2$) in the surrounding gas, significantly altering its cooling efficiency. Building on our previous work \citep{Zhang2025}, we demonstrate that massive PBHs enhance local overdensities, expediting the formation of DM halos and attracting the surrounding gas from the IGM, while simultaneously introducing thermal feedback that delays gas cooling and collapse. In this study, we focus on conditions for DCBH formation around a fiducial heating efficiency value of $\epsilon_r \sim 0.5\%$, previously identified by \cite{Zhang2025} as critical to the formation of collapsing gas clouds in the absence of LW radiation.

Another vital condition explored in our study is the baryon-DM relative streaming velocity \citep[e.g.,][]{Stacy2011, Schauer2017MNRAS.471.4878S, Schauer2019MNRAS.484.3510S}. Unlike the standard $\Lambda$CDM scenario, where streaming velocities typically delay the formation of collapsing gas clouds, our simulations suggest that streaming effects can enhance collapse by creating dense pockets of gas behind the PBH trajectory through the IGM. This effect becomes particularly significant at higher initial streaming velocities ($v_{\rm b\chi} \gtrsim 0.8 \sigma_{\rm b\chi}$), emphasizing the critical role of streaming velocities in promoting the formation of massive pristine gas clouds. Conversely, as demonstrated in our \texttt{PBH\_LW\_wstr\_fd005} and \texttt{PBH\_LW\_fd005} runs, lowering the initial streaming velocity reduces the offset between the PBH and the gas cloud's center of mass, causing dense gas during the initial collapse phase to be rapidly engulfed by the central black hole. 

An illustrative example of the gas collapse process is presented in Figure~\ref{fig:SlicePlot}, capturing the onset of collapse from the \texttt{PBH\_LW\_str\_fd005} simulation at $z \simeq 17.4$. Similar to~\cite{Zhang2025}, cold gas flows in from the IGM at speeds of $\sim 50~\rm km$\,s$^{-1}$, competing against the hot gas outflows due to BH accretion and forming a dense gas cocoon around the PBH. The collapse occurs within a dense pocket, characterized by a compact core radius of $\lesssim \mathcal{O}(1)$\,pc. 
At $\sim$10\,pc from the PBH, the collapsing gas cloud reaches a hydrogen number density of $n_{\rm H}\sim 10^6~\rm cm^{-3}$ and a temperature of $T \sim 10^4~$K, while the central PBH accretes at roughly $1\%$ of its Eddington rate. Using these parameters in Equ.~(\ref{eq:jlw}), we calculate a LW radiation intensity of $J_{\rm LW}/J_{21} \sim 4\times 10^5$, which greatly exceeds the critical threshold of $J_{\rm crit} \sim 10-10^3$ necessary for H$_2$ suppression \citep[e.g.,][]{Sugimura2014MNRAS.445..544S,Agarwal2016MNRAS.459.4209A, Trinca2022MNRAS.511..616T}, implying dominant atomic hydrogen cooling\footnote{The physics behind this critical value is rather complex \citep[e.g.,][]{Woods2019}, so instead we show a range of values here from the literature.}. A similar gas configuration is also observed in the \texttt{PBH\_LW\_sstr\_fd005} run at $z\sim 15$ and in the \texttt{PBH\_LW\_mstr\_fd005} run at $z\sim 11.7$, with collapsing particles similarly positioned relative to the PBH. Although variations in initial relative streaming velocities were tested, no clear correlation emerged between streaming velocity amplitudes and collapse initiation times. This implies a stochastic nature of DCBH formation around PBHs, and future studies should consider multiple random realizations for each streaming velocity to evaluate the statistics and possible trends. 

Figure~\ref{fig:Tn} provides further details on the conditions necessary for gas cloud collapse under LW radiation through phase diagrams depicting temperature ($T$) versus hydrogen number density ($n_{\mathrm{H}}$) when the first collapsing particles were identified. These diagrams display several simulation settings\footnote{Note that the results for the \texttt{CDM} and \texttt{PBH\_fd005} runs are reproduced from \cite{Zhang2025}, where LW radiation backgrounds were absent and the collapse of gas was governed by a Jeans instability criterion.}: the baseline \texttt{CDM} simulation at $z = 14.9$, the \texttt{PBH\_fd005} run without LW radiation at $z = 29.6$, the \texttt{PBH\_LW\_fd005} run at $z = 10.6$, and the \texttt{PBH\_LW\_str\_fd005} run at $z = 17.4$. A constant feedback efficiency of $\epsilon_r = 0.5\%$ is assumed for simulations containing a PBH. These phase diagrams illustrate substantial variations in cooling pathways, especially a shift from molecular hydrogen cooling (in the \texttt{CDM} and \texttt{PBH\_fd005} runs) to atomic cooling (in the \texttt{PBH\_LW\_fd005} and \texttt{PBH\_LW\_str\_fd005} runs) under LW radiation. The bifurcation in the \texttt{PBH\_LW\_fd005} case is reflecting the collapse of pristine gas through $\rm H_2$ cooling within halos $\sim$15\,kpc away from the central PBH where $J_{\rm LW} < J_{\rm crit}$, such that the thermal evolution there is not significantly affected by LW radiation. In this case, moreover, the gas surrounding the PBH is cooling via atomic hydrogen, thus exhibiting a nearly isothermal trend. However, this gas cloud is never able to reach densities of $n_{\rm H} \gtrsim 10^4~\rm cm^{-3}$ before being accreted or evaporated by the BH, thus avoiding runaway collapse until the end of the simulation. Notably, the \texttt{PBH\_LW\_str\_fd005} simulation highlights the key role of (sufficiently strong) baryon-DM streaming in triggering the collapse of dense, quasi-isothermal gas clouds ($\sim 10^5~\Msun$, $T\sim 8000-9000~$K, and $n_{\rm H}\gtrsim 10^4~\rm cm^{-3}$).

\subsection{Mass Inflow and DCBH Formation Criterion}\label{subsec:inflowDCBH}
\begin{figure*}[ht!]
\centering
    \includegraphics[width=\linewidth]{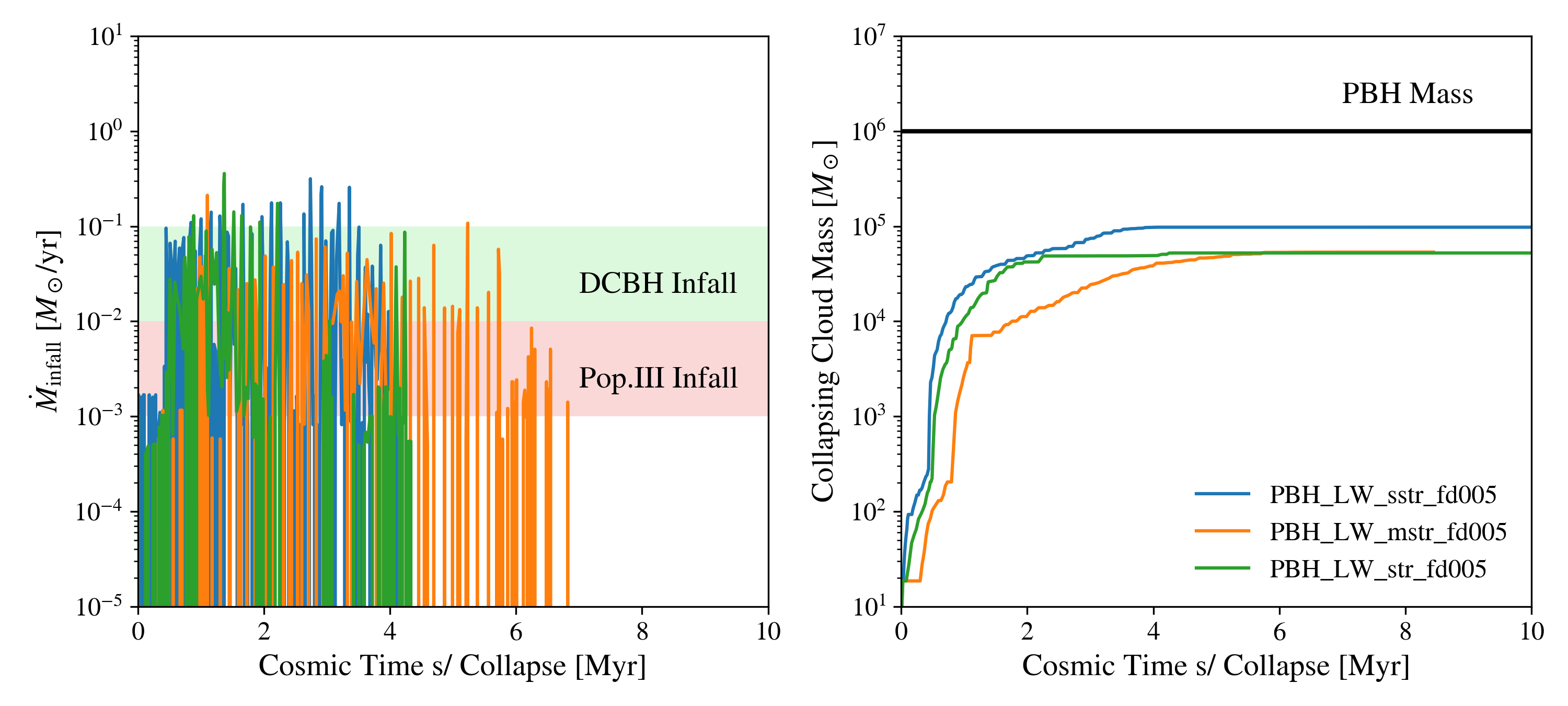}
    \vspace{1ex}
    \caption{Evolution of gas infall rate (\textbf{left panel}) and total collapsing cloud mass (\textbf{right panel}) as a function of cosmic time since the initial collapse event, comparing select simulation scenarios: \texttt{PBH\_LW\_sstr\_fd005} (blue), \texttt{PBH\_LW\_mstr\_fd005} (orange) and \texttt{PBH\_LW\_str\_fd005} (green). The left panel compares our simulation results with typical inflow rates encountered in Pop~III star formation ($\sim 10^{-2} - 10^{-3} ~\Msun$\,yr$^{-1}$; light-red shading) and the critical rate required for DCBH formation ($\sim 0.01 - 0.1~ \Msun$\,yr$^{-1}$; light-green shading). The right panel summarizes the growth of the collapsing cloud mass within $\sim 5~\rm Myr$ of cosmic time, approaching values characteristic of DCBH seeds ($\sim 10^5 ~\Msun$), comparing to the primary PBH mass ($\sim 10^6~\Msun$; solid line). As can be seen, cases that combine LW feedback with the presence of (strong) baryon-DM streaming encounter conditions favorable for massive black hole seed formation.}
    \label{fig:inflow}
\end{figure*}

Although our simulations identify dense, gravitationally collapsing atomic-cooling gas clouds near PBHs, we emphasize that we do not directly simulate the formation of a DCBH. Instead, we identify conditions conducive to the formation of a SMS as DCBH progenitor \citep[e.g.,][]{Liu2024_Mass}. In this section, to further elucidate the evolution of the collapsing gas cloud and the conditions favorable for DCBH formation, we closely analyze the temporal progression of the infall rate and mass accumulation within the collapsing region.

As demonstrated in Figure~\ref{fig:inflow}, simulation cases including the effect of streaming with $v_{\rm b\chi} \gtrsim 0.8 \sigma_{\rm b\chi}$ (\texttt{PBH\_LW\_sstr\_fd005}, \texttt{PBH\_LW\_mstr\_fd005} and \texttt{PBH\_LW\_str\_fd005}) result in the formation of massive collapsing clouds $\sim$6~Myr after the initial collapse. Throughout this process, the gas infall rate consistently exceeds $\gtrsim 10^{-3}~\Msun$\,yr$^{-1}$, with peak values frequently surpassing $\gtrsim 10^{-2}~\Msun$\,yr$^{-1}$. In our simulations, since accretion onto sink particles was not implemented and the resolution limit does not extend to the length scale of a proto-star, we approximate the gas infall  rate onto the proto-star by differentiating the total collapsing particle masses with respect to time.

To interpret these inflow rates physically, we compare them with characteristic values associated with both Pop~III star formation and DCBH formation (see also \citealt{Liu2024_Mass}). In zeroth order, the gas infall rate for a self-gravitating spherical gas cloud can be estimated by dividing the Jeans mass by the free-fall timescale \citep[e.g.,][]{Stahler1980}:
\begin{equation}
\dot{M}_{\rm infall} \sim \frac{M_{\rm J}}{t_{\rm ff}} \simeq \frac{c^3_s}{G} \simeq 4\times 10^{-3} \left(\frac{T_{\rm gas}}{10^3~\rm K}\right)^{3/2}\Msun\rm\,yr^{-1} .\label{eq:infall}
\end{equation}
For gas cooled predominantly via atomic hydrogen with temperatures of $T_{\rm gas}\sim 5000 - 10^4~\rm K$, the gas inflow rates naturally exceed the critical rate required for DCBH formation via (feedback-free, bloated) SMSs $\sim 0.01-0.1~\Msun$\,yr$^{-1}$ \citep[see e.g.,][]{Haemmerle2018,Inayoshi:2020,Herrington2023,Liu2024_Mass}. Conversely, gas clouds undergoing molecular cooling typically exhibit temperatures of $\sim 400-1000~\rm K$, corresponding to a typical inflow rate of $\sim 10^{-2} - 10^{-3}~\Msun$\,yr$^{-1}$ for conventional Pop~III star formation scenarios \citep[e.g.,][]{Bromm2013,Klessen:2023FirstStars}. 

In our simulations, we find that the compact core grows rapidly to $\sim 5\times10^4 - 10^5~\Msun$ within $\sim 6$ Myr, consistent with theoretical expectations for DCBH seed masses \citep[e.g.,][]{Becerra2018_DCBH,Latif2022_DCBH,Chon2025}. The sustained elevated inflow rate thus strongly supports rapid mass accumulation onto the core, indicating a clear pathway for the formation of a supermassive object in the center of the cloud. Comparing the mass of this collapsed core with that of the central PBH yields a mass ratio of $q \sim 0.05$–0.1, confirming the emergence of a massive binary system.

Here, we have assumed that the inflow feeds a single protostar. The simulated inflow rates frequently exceed the critical threshold for entering the bloated phase, in which the protostar develops an extended, cool surface \citep[e.g.,][]{Hosokawa2013}. In this regime, ionizing feedback is suppressed, enabling sustained accretion and ultimately leading to the formation of a SMS with mass $\sim 10^4$–$10^5\, M_\odot$ \citep[e.g.,][]{Toyouchi2023}. Previous simulations further demonstrate that such high inflow rates can sustain SMS growth \textit{even in the presence of moderate fragmentation}, through mechanisms such as super-competitive accretion and accretion-aided stellar collisions \citep[e.g.,][]{Chon2020SMS,Chon2025,Reinoso2023,Solar2025}. Alternatively, if strong fragmentation is able to significantly reduce the accretion rate of individual protostars, thus suppressing these mechanisms and causing the so-called fragmentation-induced starvation \citep{Prole2022}, a dense star cluster will form \citep[e.g.,][]{Sakurai2017,Hirano2018ApJ...855...17H,Wang2022}. The core collapse of such dense clusters can still produce massive stars/black holes above $1000\, M_\odot$ through runaway collisions.

Although our simulations resolve gas collapse on scales of $\gtrsim1–10\,$pc, they do not resolve sub-pc processes such as turbulence, radiative transfer, and angular momentum transport, which are critical for determining whether the gas inflow directly feeds a single central object or instead settles into a rotationally supported disk or fragments into multiple cores \citep{Latif2020ApJSMBHB,Latif2022_DCBH,Regan2024}. While we track collapsing gas clouds using a sink-particle formalism based on Jeans instability, this prescription cannot fully capture the complex thermodynamic and dynamical evolution at smaller scales. 
Higher-resolution simulations incorporating full radiative (magneto-)hydrodynamics \citep[e.g.,][]{Regan2020,Latif2021,Latif2022_DCBH,Regan2024} 
are essential to determine the fate of such collapsing gas clouds and the robustness of our binary SMBH formation channel.

\section{Empirical Signatures} \label{sec:implications}

The SMBH binary formation scenario explored in this study has significant implications for current and future observational efforts across both the electromagnetic (EM) and gravitational wave (GW) domains, providing a promising pathway for multi-messenger astrophysics \citep[see e.g.,][]{DeRosaSMBHB2019NewAR..8601525D}. Detecting such signals would enhance our understanding of SMBH binary formation mechanisms, accretion processes, and mass evolution, directly linking theoretical predictions from early-Universe scenarios to observable gravitational wave phenomena.

SMBH binaries originating from this PBH-induced DCBH pathway are initially expected to exhibit mass ratios of $q \sim \mathcal{O}(0.1)$ and separations of $\sim 10$ pc. The secondary black hole, initially embedded in a dense massive gas cloud ($n_{\rm H} \gtrsim 10^4~\rm cm^{-3}$, $M_{\rm cloud} \sim \mathcal{O}(10^5)~\Msun$) will likely experience enhanced accretion greatly exceeding the Eddington limit (i.e., for a $10^5~\Msun$ BH seed enshrouded by this dense gas cloud, the accretion rate given by the Bondi-Hoyle formula Equ.~(\ref{eq:bondi}) attains a Eddington ratio as high as $\eta \sim 100$). At this stage, the internal accretion feedback from the secondary BH is not able to stop the accretion flow~\citep[see e.g.,][]{Toyouchi2021IMBH}, and the BH can grow rapidly at a very short time scale before the external feedback from the PBH destroys the cloud, which quickly drives the mass ratio closer to unity. This also implies that the secondary BH will become a more luminous source than the primary PBH during this phase, as the PBH only accretes at $\eta \sim 0.1-0.01$. If dynamical friction between the newly formed DCBH and gas effectively facilitates orbital decay, these systems may evolve into tightly bound binaries with separations $\lesssim 1~\rm pc$ on relatively short timescales~\citep[see e.g.,][]{Fiacconi2013ApJ...777L..14F}.

After 
the PBH-DCBH binary becomes tightly bound, this scenario could offer a compelling explanation for enigmatic high-redshift sources, such as the LRDs~\citep[e.g.,][]{MattheeLRD2024ApJ...963..129M}. The characteristic V-shaped spectra of these objects could arise from dual thermal emissions—hotter mini-disks around each SMBH and a colder circumbinary disk—according to predictions from binary accretion models \citep[e.g.,][]{InayoshiBinaryBH2025arXiv}.


In addition, our findings have strong implications for gravitational wave astrophysics. Once in the GW-driven regime with a separation of $\sim 10^{-2} - 10^{-3}$\,pc, PBH-DCBH binaries become prime targets for space-based interferometers such as LISA and TianQin, which are sensitive to SMBH coalescence within the $10^3$--$10^6~\Msun$ range at millihertz frequencies \citep[e.g.,][]{TianQin2016CQGra..33c5010L, LISA2017arXiv170200786A, TianQin2025RPPh...88e6901L}. The extended inspiral phases of these binaries produce long-wavelength GW signals \citep[e.g.,][]{InayoshiSMBHGW2018ApJ, SasakiPBH2018CQGra..35f3001S}, and may also contribute to the stochastic gravitational wave background detectable by PTAs, including NANOGrav, EPTA, and CPTA \citep[e.g.,][]{Nanograv2023ApJ, PPTA2023ApJ, EPTA2023A&A}. 

While our baseline observational predictions assume the formation of a single DCBH, alternative outcomes—such as enhanced fragmentation and the formation of dense stellar clusters—may yield distinct multi-messenger signatures. 
These clusters can host a diverse black hole population ranging from ordinary stellar remnants $\sim 10-100\, M_\odot$ to intermediate-mass black holes (IMBHs $\sim 100-10^5\, M_\odot$). 
Gravitational interactions among these black holes—or with the central PBH—can produce extreme-mass-ratio inspirals, detectable by LISA and TianQin \citep[e.g.,][]{Naoz2023ApJ...955L..27N}. The interactions between black holes and stars in these systems can lead to tidal disruption events 
\citep[e.g.,][]{Inayoshi2024ApJ...966..164I, Wang2025TDE250418144W}. Moreover, dynamical processes in these dense star clusters can produce (hierarchical) mergers of IMBH binaries \citep[e.g.,][]{Wang2022,Liu2024sc}, potentially contributing to the IMBH merger event rate \citep[e.g.,][]{Fragione2022ApJ...933..170F}. Together, the complex interplay between fragmentation, feedback, and collapse mechanisms strongly motivates combining high-resolution radiation-hydrodynamic simulations with upcoming JWST, ALMA, and LISA data to robustly constrain the nature and fate of the earliest black hole seeds.

\section{Summary and Conclusions}\label{sec:conclusion}
In this study, we have presented a novel pathway for the formation of SMBH binaries through the direct-collapse mechanism induced by a primary PBH. This PBH-induced DCBH formation channel provides a natural \textit{in situ} mechanism for the emergence of SMBH binaries in the early Universe. 

The interplay between PBH gravitational seeding, accretion-driven thermal feedback, and baryon–dark matter relative streaming motions creates favorable conditions for the accumulation and collapse of massive pristine gas clouds, as was already suggested in previous work~\citep{Boyuan2022MNRAS.514.2376L,Boyuan2023arXiv231204085L,Zhang2025}. Our hydrodynamical simulations demonstrate that LW radiation, locally generated by PBH accretion, significantly alters the thermal and chemical evolution of the nearby gas, suppressing molecular hydrogen cooling and shifting the dominant cooling channel to atomic hydrogen. When combined with sufficiently high initial streaming velocities ($v_{\rm b\chi} \gtrsim 0.8 \sigma_{\rm b\chi}$), we identify a critical regime that enables the formation of gravitationally unstable, atomically-cooling gas clouds in the wake of the PBH trajectory.

During the collapse phase, our simulations show that gas inflow rates consistently exceed the threshold for DCBH formation, with $\dot{M}_{\rm infall} \gtrsim 0.01-0.1~\Msun$\,yr$^{-1}$. This leads to the inevitable and rapid formation of compact massive cores within the mass range of $\sim 5\times10^4 - 10^5~\Msun$ in $\sim 5~\rm Myr$. The resulting SMBH binaries exhibit initial mass ratios of $q \sim \mathcal{O}(0.1)$ and separations of $\sim 10$ pc, naturally evolving into gravitationally bound systems.

These PBH-DCBH binary systems are also promising sources of gravitational waves, potentially detectable with next-generation observatories such as LISA and TianQin~\citep[e.g.,][]{TianQin2016CQGra..33c5010L, LISA2017arXiv170200786A, TianQin2025RPPh...88e6901L}. Furthermore, they may serve as plausible progenitors of binary SMBHs that are invoked to explain the distinct spectral features observed in select high-$z$ JWST/ALMA targets, such as a subset of LRDs~\citep[e.g.,][]{LabbeLRD2025ApJ...978...92L}. Future work should explore a broader range of PBH masses and spatial distributions, investigating the long-term dynamical evolution of PBH-DCBH binaries in realistic cosmological environments, and consider the multi-frequency radiative transfer around these sources to derive their detailed observational signatures.  Altogether, these findings establish a fundamental connection between early-Universe black hole physics, multi-messenger astrophysics, and the broader study of dark matter. 

\begin{acknowledgments}
BL gratefully acknowledges the funding of the Royal Society University Research Fellowship and the Deutsche Forschungsgemeinschaft (DFG, German Research Foundation) under Germany's Excellence Strategy EXC 2181/1 - 390900948 (the Heidelberg STRUCTURES Excellence Cluster). The authors acknowledge the Texas Advanced Computing Center (TACC) for providing HPC resources under allocation AST23026.
\end{acknowledgments}





%

\vspace{5mm}

\software{astropy \citep{2013A&A...558A..33A,2018AJ....156..123A, 2022ApJ...935..167A},            Colossus \citep{Diemer2018ApJCOLOSSUS}         }




\bibliography{Main}{}

\begin{thebibliography}{}
\expandafter\ifx\csname natexlab\endcsname\relax\def\natexlab#1{#1}\fi
\providecommand{\url}[1]{\href{#1}{#1}}
\providecommand{\dodoi}[1]{doi:~\href{http://doi.org/#1}{\nolinkurl{#1}}}
\providecommand{\doeprint}[1]{\href{http://ascl.net/#1}{\nolinkurl{http://ascl.net/#1}}}
\providecommand{\doarXiv}[1]{\href{https://arxiv.org/abs/#1}{\nolinkurl{https://arxiv.org/abs/#1}}}

\bibitem[{T. {Abel} {et~al.}(1997){Abel}, {Anninos}, {Zhang}, \& {Norman}}]{Abel1997NewA....2..181A}
{Abel}, T., {Anninos}, P., {Zhang}, Y., \& {Norman}, M.~L. 1997, \bibinfo{title}{{Modeling primordial gas in numerical cosmology},} \na, 2, 181, \dodoi{10.1016/S1384-1076(97)00010-9}

\bibitem[{N. {Afshordi} {et~al.}(2003){Afshordi}, {McDonald}, \& {Spergel}}]{Afshordi2003ApJ...594L..71A}
{Afshordi}, N., {McDonald}, P., \& {Spergel}, D.~N. 2003, \bibinfo{title}{{Primordial Black Holes as Dark Matter: The Power Spectrum and Evaporation of Early Structures},} \apjl, 594, L71, \dodoi{10.1086/378763}

\bibitem[{B. {Agarwal} {et~al.}(2016){Agarwal}, {Smith}, {Glover}, {Natarajan}, \& {Khochfar}}]{Agarwal2016MNRAS.459.4209A}
{Agarwal}, B., {Smith}, B., {Glover}, S., {Natarajan}, P., \& {Khochfar}, S. 2016, \bibinfo{title}{{New constraints on direct collapse black hole formation in the early Universe},} \mnras, 459, 4209, \dodoi{10.1093/mnras/stw929}

\bibitem[{G. {Agazie} {et~al.}(2023){Agazie}, {Anumarlapudi}, {Archibald}, {Arzoumanian}, {Baker}, {B{\'e}csy}, {Blecha}, {Brazier}, {Brook}, {Burke-Spolaor}, {Case}, {Casey-Clyde}, {Charisi}, {Chatterjee}, {Cohen}, {Cordes}, {Cornish}, {Crawford}, {Cromartie}, {Crowter}, {Decesar}, {Demorest}, {Digman}, {Dolch}, {Drachler}, {Ferrara}, {Fiore}, {Fonseca}, {Freedman}, {Garver-Daniels}, {Gentile}, {Glaser}, {Good}, {G{\"u}ltekin}, {Hazboun}, {Hourihane}, {Jennings}, {Johnson}, {Jones}, {Kaiser}, {Kaplan}, {Kelley}, {Kerr}, {Key}, {Laal}, {Lam}, {Lamb}, {Lazio}, {Lewandowska}, {Liu}, {Lorimer}, {Luo}, {Lynch}, {Ma}, {Madison}, {McEwen}, {McKee}, {McLaughlin}, {McMann}, {Meyers}, {Meyers}, {Mingarelli}, {Mitridate}, {Ng}, {Nice}, {Ocker}, {Olum}, {Pennucci}, {Perera}, {Petrov}, {Pol}, {Radovan}, {Ransom}, {Ray}, {Romano}, {Sardesai}, {Schmiedekamp}, {Schmiedekamp}, {Schmitz}, {Shapiro-Albert}, {Siemens}, {Simon}, {Siwek}, {Stairs}, {Stinebring}, {Stovall}, {Susobhanan}, {Swiggum}, {Taylor}, {Taylor}, {Turner},
  {Unal}, {Vallisneri}, {van Haasteren}, {Vigeland}, {Wahl}, {Witt}, {Young}, \& {Nanograv Collaboration}}]{Nanograv2023ApJ}
{Agazie}, G., {Anumarlapudi}, A., {Archibald}, A.~M., {et~al.} 2023, \bibinfo{title}{{The NANOGrav 15 yr Data Set: Bayesian Limits on Gravitational Waves from Individual Supermassive Black Hole Binaries},} \apjl, 951, L50, \dodoi{10.3847/2041-8213/ace18a}

\bibitem[{Y. {Ali-Ha{\"\i}moud} \& M. {Kamionkowski}(2017){Ali-Ha{\"\i}moud} \& {Kamionkowski}}]{Ali-Haimoud2017PhRvD}
{Ali-Ha{\"\i}moud}, Y., \& {Kamionkowski}, M. 2017, \bibinfo{title}{{Cosmic microwave background limits on accreting primordial black holes},} \prd, 95, 043534, \dodoi{10.1103/PhysRevD.95.043534}

\bibitem[{P. {Amaro-Seoane} {et~al.}(2017){Amaro-Seoane}, {Audley}, {Babak}, {Baker}, {Barausse}, {Bender}, {Berti}, {Binetruy}, {Born}, {Bortoluzzi}, {Camp}, {Caprini}, {Cardoso}, {Colpi}, {Conklin}, {Cornish}, {Cutler}, {Danzmann}, {Dolesi}, {Ferraioli}, {Ferroni}, {Fitzsimons}, {Gair}, {Gesa Bote}, {Giardini}, {Gibert}, {Grimani}, {Halloin}, {Heinzel}, {Hertog}, {Hewitson}, {Holley-Bockelmann}, {Hollington}, {Hueller}, {Inchauspe}, {Jetzer}, {Karnesis}, {Killow}, {Klein}, {Klipstein}, {Korsakova}, {Larson}, {Livas}, {Lloro}, {Man}, {Mance}, {Martino}, {Mateos}, {McKenzie}, {McWilliams}, {Miller}, {Mueller}, {Nardini}, {Nelemans}, {Nofrarias}, {Petiteau}, {Pivato}, {Plagnol}, {Porter}, {Reiche}, {Robertson}, {Robertson}, {Rossi}, {Russano}, {Schutz}, {Sesana}, {Shoemaker}, {Slutsky}, {Sopuerta}, {Sumner}, {Tamanini}, {Thorpe}, {Troebs}, {Vallisneri}, {Vecchio}, {Vetrugno}, {Vitale}, {Volonteri}, {Wanner}, {Ward}, {Wass}, {Weber}, {Ziemer}, \& {Zweifel}}]{LISA2017arXiv170200786A}
{Amaro-Seoane}, P., {Audley}, H., {Babak}, S., {et~al.} 2017, \bibinfo{title}{{Laser Interferometer Space Antenna},} arXiv e-prints, arXiv:1702.00786, \dodoi{10.48550/arXiv.1702.00786}

\bibitem[{ {Astropy Collaboration} {et~al.}(2013){Astropy Collaboration}, {Robitaille}, {Tollerud}, {Greenfield}, {Droettboom}, {Bray}, {Aldcroft}, {Davis}, {Ginsburg}, {Price-Whelan}, {Kerzendorf}, {Conley}, {Crighton}, {Barbary}, {Muna}, {Ferguson}, {Grollier}, {Parikh}, {Nair}, {Unther}, {Deil}, {Woillez}, {Conseil}, {Kramer}, {Turner}, {Singer}, {Fox}, {Weaver}, {Zabalza}, {Edwards}, {Azalee Bostroem}, {Burke}, {Casey}, {Crawford}, {Dencheva}, {Ely}, {Jenness}, {Labrie}, {Lim}, {Pierfederici}, {Pontzen}, {Ptak}, {Refsdal}, {Servillat}, \& {Streicher}}]{2013A&A...558A..33A}
{Astropy Collaboration}, {Robitaille}, T.~P., {Tollerud}, E.~J., {et~al.} 2013, \bibinfo{title}{{Astropy: A community Python package for astronomy},} \aap, 558, A33, \dodoi{10.1051/0004-6361/201322068}

\bibitem[{ {Astropy Collaboration} {et~al.}(2018){Astropy Collaboration}, {Price-Whelan}, {Sip{\H{o}}cz}, {G{\"u}nther}, {Lim}, {Crawford}, {Conseil}, {Shupe}, {Craig}, {Dencheva}, {Ginsburg}, {VanderPlas}, {Bradley}, {P{\'e}rez-Su{\'a}rez}, {de Val-Borro}, {Aldcroft}, {Cruz}, {Robitaille}, {Tollerud}, {Ardelean}, {Babej}, {Bach}, {Bachetti}, {Bakanov}, {Bamford}, {Barentsen}, {Barmby}, {Baumbach}, {Berry}, {Biscani}, {Boquien}, {Bostroem}, {Bouma}, {Brammer}, {Bray}, {Breytenbach}, {Buddelmeijer}, {Burke}, {Calderone}, {Cano Rodr{\'\i}guez}, {Cara}, {Cardoso}, {Cheedella}, {Copin}, {Corrales}, {Crichton}, {D'Avella}, {Deil}, {Depagne}, {Dietrich}, {Donath}, {Droettboom}, {Earl}, {Erben}, {Fabbro}, {Ferreira}, {Finethy}, {Fox}, {Garrison}, {Gibbons}, {Goldstein}, {Gommers}, {Greco}, {Greenfield}, {Groener}, {Grollier}, {Hagen}, {Hirst}, {Homeier}, {Horton}, {Hosseinzadeh}, {Hu}, {Hunkeler}, {Ivezi{\'c}}, {Jain}, {Jenness}, {Kanarek}, {Kendrew}, {Kern}, {Kerzendorf}, {Khvalko}, {King}, {Kirkby}, {Kulkarni},
  {Kumar}, {Lee}, {Lenz}, {Littlefair}, {Ma}, {Macleod}, {Mastropietro}, {McCully}, {Montagnac}, {Morris}, {Mueller}, {Mumford}, {Muna}, {Murphy}, {Nelson}, {Nguyen}, {Ninan}, {N{\"o}the}, {Ogaz}, {Oh}, {Parejko}, {Parley}, {Pascual}, {Patil}, {Patil}, {Plunkett}, {Prochaska}, {Rastogi}, {Reddy Janga}, {Sabater}, {Sakurikar}, {Seifert}, {Sherbert}, {Sherwood-Taylor}, {Shih}, {Sick}, {Silbiger}, {Singanamalla}, {Singer}, {Sladen}, {Sooley}, {Sornarajah}, {Streicher}, {Teuben}, {Thomas}, {Tremblay}, {Turner}, {Terr{\'o}n}, {van Kerkwijk}, {de la Vega}, {Watkins}, {Weaver}, {Whitmore}, {Woillez}, {Zabalza}, \& {Astropy Contributors}}]{2018AJ....156..123A}
{Astropy Collaboration}, {Price-Whelan}, A.~M., {Sip{\H{o}}cz}, B.~M., {et~al.} 2018, \bibinfo{title}{{The Astropy Project: Building an Open-science Project and Status of the v2.0 Core Package},} \aj, 156, 123, \dodoi{10.3847/1538-3881/aabc4f}

\bibitem[{ {Astropy Collaboration} {et~al.}(2022){Astropy Collaboration}, {Price-Whelan}, {Lim}, {Earl}, {Starkman}, {Bradley}, {Shupe}, {Patil}, {Corrales}, {Brasseur}, {N{\"o}the}, {Donath}, {Tollerud}, {Morris}, {Ginsburg}, {Vaher}, {Weaver}, {Tocknell}, {Jamieson}, {van Kerkwijk}, {Robitaille}, {Merry}, {Bachetti}, {G{\"u}nther}, {Aldcroft}, {Alvarado-Montes}, {Archibald}, {B{\'o}di}, {Bapat}, {Barentsen}, {Baz{\'a}n}, {Biswas}, {Boquien}, {Burke}, {Cara}, {Cara}, {Conroy}, {Conseil}, {Craig}, {Cross}, {Cruz}, {D'Eugenio}, {Dencheva}, {Devillepoix}, {Dietrich}, {Eigenbrot}, {Erben}, {Ferreira}, {Foreman-Mackey}, {Fox}, {Freij}, {Garg}, {Geda}, {Glattly}, {Gondhalekar}, {Gordon}, {Grant}, {Greenfield}, {Groener}, {Guest}, {Gurovich}, {Handberg}, {Hart}, {Hatfield-Dodds}, {Homeier}, {Hosseinzadeh}, {Jenness}, {Jones}, {Joseph}, {Kalmbach}, {Karamehmetoglu}, {Ka{\l}uszy{\'n}ski}, {Kelley}, {Kern}, {Kerzendorf}, {Koch}, {Kulumani}, {Lee}, {Ly}, {Ma}, {MacBride}, {Maljaars}, {Muna}, {Murphy}, {Norman},
  {O'Steen}, {Oman}, {Pacifici}, {Pascual}, {Pascual-Granado}, {Patil}, {Perren}, {Pickering}, {Rastogi}, {Roulston}, {Ryan}, {Rykoff}, {Sabater}, {Sakurikar}, {Salgado}, {Sanghi}, {Saunders}, {Savchenko}, {Schwardt}, {Seifert-Eckert}, {Shih}, {Jain}, {Shukla}, {Sick}, {Simpson}, {Singanamalla}, {Singer}, {Singhal}, {Sinha}, {Sip{\H{o}}cz}, {Spitler}, {Stansby}, {Streicher}, {{\v{S}}umak}, {Swinbank}, {Taranu}, {Tewary}, {Tremblay}, {de Val-Borro}, {Van Kooten}, {Vasovi{\'c}}, {Verma}, {de Miranda Cardoso}, {Williams}, {Wilson}, {Winkel}, {Wood-Vasey}, {Xue}, {Yoachim}, {Zhang}, {Zonca}, \& {Astropy Project Contributors}}]{2022ApJ...935..167A}
{Astropy Collaboration}, {Price-Whelan}, A.~M., {Lim}, P.~L., {et~al.} 2022, \bibinfo{title}{{The Astropy Project: Sustaining and Growing a Community-oriented Open-source Project and the Latest Major Release (v5.0) of the Core Package},} \apj, 935, 167, \dodoi{10.3847/1538-4357/ac7c74}

\bibitem[{F. {Becerra} {et~al.}(2018{\natexlab{a}}){Becerra}, {Marinacci}, {Bromm}, \& {Hernquist}}]{Becerra2018_DCBH}
{Becerra}, F., {Marinacci}, F., {Bromm}, V., \& {Hernquist}, L.~E. 2018{\natexlab{a}}, \bibinfo{title}{{Assembly of supermassive black hole seeds},} \mnras, 480, 5029, \dodoi{10.1093/mnras/sty2210}

\bibitem[{F. {Becerra} {et~al.}(2018{\natexlab{b}}){Becerra}, {Marinacci}, {Inayoshi}, {Bromm}, \& {Hernquist}}]{Becerra2018_proto}
{Becerra}, F., {Marinacci}, F., {Inayoshi}, K., {Bromm}, V., \& {Hernquist}, L.~E. 2018{\natexlab{b}}, \bibinfo{title}{{Opacity Limit for Supermassive Protostars},} \apj, 857, 138, \dodoi{10.3847/1538-4357/aab8f4}

\bibitem[{M.~C. {Begelman} {et~al.}(1980){Begelman}, {Blandford}, \& {Rees}}]{BegelmanSMBHB1980Nature}
{Begelman}, M.~C., {Blandford}, R.~D., \& {Rees}, M.~J. 1980, \bibinfo{title}{{Massive black hole binaries in active galactic nuclei},} \nat, 287, 307, \dodoi{10.1038/287307a0}

\bibitem[{M.~C. {Begelman} {et~al.}(2006){Begelman}, {Volonteri}, \& {Rees}}]{Begelman2006:DCBH}
{Begelman}, M.~C., {Volonteri}, M., \& {Rees}, M.~J. 2006, \bibinfo{title}{{Formation of supermassive black holes by direct collapse in pre-galactic haloes},} \mnras, 370, 289, \dodoi{10.1111/j.1365-2966.2006.10467.x}

\bibitem[{K.~M. {Belotsky} {et~al.}(2019){Belotsky}, {Dokuchaev}, {Eroshenko}, {Esipova}, {Khlopov}, {Khromykh}, {Kirillov}, {Nikulin}, {Rubin}, \& {Svadkovsky}}]{Belotsky2019}
{Belotsky}, K.~M., {Dokuchaev}, V.~I., {Eroshenko}, Y.~N., {et~al.} 2019, \bibinfo{title}{{Clusters of Primordial Black Holes},} Eur. Phys. J. C, 79, 246, \dodoi{10.1140/epjc/s10052-019-6741-4}

\bibitem[{{\'A}. {Bogd{\'a}n} {et~al.}(2024){Bogd{\'a}n}, {Goulding}, {Natarajan}, {Kov{\'a}cs}, {Tremblay}, {Chadayammuri}, {Volonteri}, {Kraft}, {Forman}, {Jones}, {Churazov}, \& {Zhuravleva}}]{Bogdan:2023UHZ1}
{Bogd{\'a}n}, {\'A}., {Goulding}, A.~D., {Natarajan}, P., {et~al.} 2024, \bibinfo{title}{{Evidence for heavy-seed origin of early supermassive black holes from a z {\ensuremath{\approx}} 10 X-ray quasar},} Nature Astronomy, 8, 126, \dodoi{10.1038/s41550-023-02111-9}

\bibitem[{V. {Bromm}(2013){Bromm}}]{Bromm2013}
{Bromm}, V. 2013, \bibinfo{title}{{Formation of the first stars},} Reports on Progress in Physics, 76, 112901, \dodoi{10.1088/0034-4885/76/11/112901}

\bibitem[{V. {Bromm} {et~al.}(2002){Bromm}, {Coppi}, \& {Larson}}]{Bromm2002ApJ...564...23B}
{Bromm}, V., {Coppi}, P.~S., \& {Larson}, R.~B. 2002, \bibinfo{title}{{The Formation of the First Stars. I. The Primordial Star-forming Cloud},} \apj, 564, 23, \dodoi{10.1086/323947}

\bibitem[{V. {Bromm} \& A. {Loeb}(2003){Bromm} \& {Loeb}}]{BrommDCBH2003ApJ...596...34B}
{Bromm}, V., \& {Loeb}, A. 2003, \bibinfo{title}{{Formation of the First Supermassive Black Holes},} \apj, 596, 34, \dodoi{10.1086/377529}

\bibitem[{N. {Cappelluti} {et~al.}(2022){Cappelluti}, {Hasinger}, \& {Natarajan}}]{Cappelluti2022ApJ}
{Cappelluti}, N., {Hasinger}, G., \& {Natarajan}, P. 2022, \bibinfo{title}{{Exploring the High-redshift PBH-{\ensuremath{\Lambda}}CDM Universe: Early Black Hole Seeding, the First Stars and Cosmic Radiation Backgrounds},} \apj, 926, 205, \dodoi{10.3847/1538-4357/ac332d}

\bibitem[{B. {Carr} {et~al.}(2021{\natexlab{a}}){Carr}, {Clesse}, {Garc{\'\i}a-Bellido}, \& {K{\"u}hnel}}]{Carr2021PDU....3100755C}
{Carr}, B., {Clesse}, S., {Garc{\'\i}a-Bellido}, J., \& {K{\"u}hnel}, F. 2021{\natexlab{a}}, \bibinfo{title}{{Cosmic conundra explained by thermal history and primordial black holes},} Physics of the Dark Universe, 31, 100755, \dodoi{10.1016/j.dark.2020.100755}

\bibitem[{B. {Carr} {et~al.}(2021{\natexlab{b}}){Carr}, {Kohri}, {Sendouda}, \& {Yokoyama}}]{Carr2021}
{Carr}, B., {Kohri}, K., {Sendouda}, Y., \& {Yokoyama}, J. 2021{\natexlab{b}}, \bibinfo{title}{{Constraints on primordial black holes},} Reports on Progress in Physics, 84, 116902, \dodoi{10.1088/1361-6633/ac1e31}

\bibitem[{B. {Carr} \& F. {K{\"u}hnel}(2020){Carr} \& {K{\"u}hnel}}]{Carr2020ARNPS..70..355C}
{Carr}, B., \& {K{\"u}hnel}, F. 2020, \bibinfo{title}{{Primordial Black Holes as Dark Matter: Recent Developments},} Annual Review of Nuclear and Particle Science, 70, 355, \dodoi{10.1146/annurev-nucl-050520-125911}

\bibitem[{B. {Carr} \& J. {Silk}(2018){Carr} \& {Silk}}]{Carr2018MNRAS.478.3756C}
{Carr}, B., \& {Silk}, J. 2018, \bibinfo{title}{{Primordial black holes as generators of cosmic structures},} \mnras, 478, 3756, \dodoi{10.1093/mnras/sty1204}

\bibitem[{B.~J. {Carr}(1975){Carr}}]{Carr1975ApJ...201....1C}
{Carr}, B.~J. 1975, \bibinfo{title}{{The primordial black hole mass spectrum.},} \apj, 201, 1, \dodoi{10.1086/153853}

\bibitem[{C. {Casanueva-Villarreal} {et~al.}(2025){Casanueva-Villarreal}, {Padilla}, {Tissera}, {Liu}, \& {Bromm}}]{Casanueva-Villarreal2025A&A...699A..49C}
{Casanueva-Villarreal}, C., {Padilla}, N., {Tissera}, P.~B., {Liu}, B., \& {Bromm}, V. 2025, \bibinfo{title}{{Primordial black holes as dark matter candidates: Multi-frequency constraints from cosmic radiation backgrounds},} \aap, 699, A49, \dodoi{10.1051/0004-6361/202554032}

\bibitem[{J. {Chluba} {et~al.}(2012){Chluba}, {Erickcek}, \& {Ben-Dayan}}]{Chluba2012ApJ...758...76C}
{Chluba}, J., {Erickcek}, A.~L., \& {Ben-Dayan}, I. 2012, \bibinfo{title}{{Probing the Inflaton: Small-scale Power Spectrum Constraints from Measurements of the Cosmic Microwave Background Energy Spectrum},} \apj, 758, 76, \dodoi{10.1088/0004-637X/758/2/76}

\bibitem[{J. {Chluba} {et~al.}(2021){Chluba}, {Abitbol}, {Aghanim}, {Ali-Ha{\"\i}moud}, {Alvarez}, {Basu}, {Bolliet}, {Burigana}, {de Bernardis}, {Delabrouille}, {Dimastrogiovanni}, {Finelli}, {Fixsen}, {Hart}, {Hern{\'a}ndez-Monteagudo}, {Hill}, {Kogut}, {Kohri}, {Lesgourgues}, {Maffei}, {Mather}, {Mukherjee}, {Patil}, {Ravenni}, {Remazeilles}, {Rotti}, {Rubi{\~n}o-Martin}, {Silk}, {Sunyaev}, \& {Switzer}}]{Chluba2021ExA....51.1515C}
{Chluba}, J., {Abitbol}, M.~H., {Aghanim}, N., {et~al.} 2021, \bibinfo{title}{{New horizons in cosmology with spectral distortions of the cosmic microwave background},} Experimental Astronomy, 51, 1515, \dodoi{10.1007/s10686-021-09729-5}

\bibitem[{S. {Chon} \& K. {Omukai}(2020){Chon} \& {Omukai}}]{Chon2020SMS}
{Chon}, S., \& {Omukai}, K. 2020, \bibinfo{title}{{Supermassive star formation via super competitive accretion in slightly metal-enriched clouds},} \mnras, 494, 2851, \dodoi{10.1093/mnras/staa863}

\bibitem[{S. {Chon} \& K. {Omukai}(2025){Chon} \& {Omukai}}]{Chon2025}
{Chon}, S., \& {Omukai}, K. 2025, \bibinfo{title}{{Formation of supermassive stars and dense star clusters in metal-poor clouds exposed to strong FUV radiation},} \mnras, 539, 2561, \dodoi{10.1093/mnras/staf598}

\bibitem[{P.~E. {Colazo} {et~al.}(2024){Colazo}, {Stasyszyn}, \& {Padilla}}]{Colazo2024}
{Colazo}, P.~E., {Stasyszyn}, F., \& {Padilla}, N. 2024, \bibinfo{title}{{Structure formation with primordial black holes to alleviate early star formation tension revealed by JWST},} \aap, 685, L8, \dodoi{10.1051/0004-6361/202449565}

\bibitem[{B. {Cyr}(2025){Cyr}}]{Cyr2025arXiv250717833C}
{Cyr}, B. 2025, \bibinfo{title}{{Not-quite-primordial black holes seeded by cosmic string loops},} arXiv e-prints, arXiv:2507.17833, \dodoi{10.48550/arXiv.2507.17833}

\bibitem[{P. {Dayal} \& R. {Maiolino}(2025){Dayal} \& {Maiolino}}]{DayalPBH2025arXiv250608116D}
{Dayal}, P., \& {Maiolino}, R. 2025, \bibinfo{title}{{The properties of primordially-seeded black holes and their hosts in the first billion years: implications for JWST},} arXiv e-prints, arXiv:2506.08116, \dodoi{10.48550/arXiv.2506.08116}

\bibitem[{V. {De Luca} {et~al.}(2019){De Luca}, {Desjacques}, {Franciolini}, {Malhotra}, \& {Riotto}}]{DeLuca2019}
{De Luca}, V., {Desjacques}, V., {Franciolini}, G., {Malhotra}, A., \& {Riotto}, A. 2019, \bibinfo{title}{{The initial spin probability distribution of primordial black holes},} \jcap, 2019, 018, \dodoi{10.1088/1475-7516/2019/05/018}

\bibitem[{V. {De Luca} {et~al.}(2020){De Luca}, {Franciolini}, {Pani}, \& {Riotto}}]{Deluca2020JCAP...06..044D}
{De Luca}, V., {Franciolini}, G., {Pani}, P., \& {Riotto}, A. 2020, \bibinfo{title}{{Primordial black holes confront LIGO/Virgo data: current situation},} \jcap, 2020, 044, \dodoi{10.1088/1475-7516/2020/06/044}

\bibitem[{A. {De Rosa} {et~al.}(2019){De Rosa}, {Vignali}, {Bogdanovi{\'c}}, {Capelo}, {Charisi}, {Dotti}, {Husemann}, {Lusso}, {Mayer}, {Paragi}, {Runnoe}, {Sesana}, {Steinborn}, {Bianchi}, {Colpi}, {del Valle}, {Frey}, {Gab{\'a}nyi}, {Giustini}, {Guainazzi}, {Haiman}, {Herrera Ruiz}, {Herrero-Illana}, {Iwasawa}, {Komossa}, {Lena}, {Loiseau}, {Perez-Torres}, {Piconcelli}, \& {Volonteri}}]{DeRosaSMBHB2019NewAR..8601525D}
{De Rosa}, A., {Vignali}, C., {Bogdanovi{\'c}}, T., {et~al.} 2019, \bibinfo{title}{{The quest for dual and binary supermassive black holes: A multi-messenger view},} \nar, 86, 101525, \dodoi{10.1016/j.newar.2020.101525}

\bibitem[{B. {Diemer}(2018){Diemer}}]{Diemer2018ApJCOLOSSUS}
{Diemer}, B. 2018, \bibinfo{title}{{COLOSSUS: A Python Toolkit for Cosmology, Large-scale Structure, and Dark Matter Halos},} \apjs, 239, 35, \dodoi{10.3847/1538-4365/aaee8c}

\bibitem[{B.~T. {Draine} \& F. {Bertoldi}(1996){Draine} \& {Bertoldi}}]{DB1996ApJ...468..269D}
{Draine}, B.~T., \& {Bertoldi}, F. 1996, \bibinfo{title}{{Structure of Stationary Photodissociation Fronts},} \apj, 468, 269, \dodoi{10.1086/177689}

\bibitem[{ {EPTA Collaboration} {et~al.}(2023){EPTA Collaboration}, {InPTA Collaboration}, {Antoniadis}, {Arumugam}, {Arumugam}, {Babak}, {Bagchi}, {Bak Nielsen}, {Bassa}, {Bathula}, {Berthereau}, {Bonetti}, {Bortolas}, {Brook}, {Burgay}, {Caballero}, {Chalumeau}, {Champion}, {Chanlaridis}, {Chen}, {Cognard}, {Dandapat}, {Deb}, {Desai}, {Desvignes}, {Dhanda-Batra}, {Dwivedi}, {Falxa}, {Ferdman}, {Franchini}, {Gair}, {Goncharov}, {Gopakumar}, {Graikou}, {Grie{\ss}meier}, {Guillemot}, {Guo}, {Gupta}, {Hisano}, {Hu}, {Iraci}, {Izquierdo-Villalba}, {Jang}, {Jawor}, {Janssen}, {Jessner}, {Joshi}, {Kareem}, {Karuppusamy}, {Keane}, {Keith}, {Kharbanda}, {Kikunaga}, {Kolhe}, {Kramer}, {Krishnakumar}, {Lackeos}, {Lee}, {Liu}, {Liu}, {Lyne}, {McKee}, {Maan}, {Main}, {Mickaliger}, {Ni{\c{t}}u}, {Nobleson}, {Paladi}, {Parthasarathy}, {Perera}, {Perrodin}, {Petiteau}, {Porayko}, {Possenti}, {Prabu}, {Quelquejay Leclere}, {Rana}, {Samajdar}, {Sanidas}, {Sesana}, {Shaifullah}, {Singha}, {Speri}, {Spiewak}, {Srivastava},
  {Stappers}, {Surnis}, {Susarla}, {Susobhanan}, {Takahashi}, {Tarafdar}, {Theureau}, {Tiburzi}, {van der Wateren}, {Vecchio}, {Venkatraman Krishnan}, {Verbiest}, {Wang}, {Wang}, \& {Wu}}]{EPTA2023A&A}
{EPTA Collaboration}, {InPTA Collaboration}, {Antoniadis}, J., {et~al.} 2023, \bibinfo{title}{{The second data release from the European Pulsar Timing Array. III. Search for gravitational wave signals},} \aap, 678, A50, \dodoi{10.1051/0004-6361/202346844}

\bibitem[{A. {Escriv{\`a}}(2022){Escriv{\`a}}}]{Escriva2022}
{Escriv{\`a}}, A. 2022, \bibinfo{title}{{PBH Formation from Spherically Symmetric Hydrodynamical Perturbations: A Review},} Universe, 8, 66, \dodoi{10.3390/universe8020066}

\bibitem[{D. {Fiacconi} {et~al.}(2013){Fiacconi}, {Mayer}, {Ro{\v{s}}kar}, \& {Colpi}}]{Fiacconi2013ApJ...777L..14F}
{Fiacconi}, D., {Mayer}, L., {Ro{\v{s}}kar}, R., \& {Colpi}, M. 2013, \bibinfo{title}{{Massive Black Hole Pairs in Clumpy, Self-gravitating Circumnuclear Disks: Stochastic Orbital Decay},} \apjl, 777, L14, \dodoi{10.1088/2041-8205/777/1/L14}

\bibitem[{G. {Fragione} {et~al.}(2022){Fragione}, {Loeb}, {Kocsis}, \& {Rasio}}]{Fragione2022ApJ...933..170F}
{Fragione}, G., {Loeb}, A., {Kocsis}, B., \& {Rasio}, F.~A. 2022, \bibinfo{title}{{Merger Rates of Intermediate-mass Black Hole Binaries in Nuclear Star Clusters},} \apj, 933, 170, \dodoi{10.3847/1538-4357/ac75d0}

\bibitem[{D. {Galli} \& F. {Palla}(2013){Galli} \& {Palla}}]{Galli2013ARA&A..51..163G}
{Galli}, D., \& {Palla}, F. 2013, \bibinfo{title}{{The Dawn of Chemistry},} \araa, 51, 163, \dodoi{10.1146/annurev-astro-082812-141029}

\bibitem[{A.~D. {Goulding} {et~al.}(2023){Goulding}, {Greene}, {Setton}, {Labbe}, {Bezanson}, {Miller}, {Atek}, {Bogd{\'a}n}, {Brammer}, {Chemerynska}, {Cutler}, {Dayal}, {Fudamoto}, {Fujimoto}, {Furtak}, {Kokorev}, {Khullar}, {Leja}, {Marchesini}, {Natarajan}, {Nelson}, {Oesch}, {Pan}, {Papovich}, {Price}, {van Dokkum}, {Wang}, {Weaver}, {Whitaker}, \& {Zitrin}}]{Goulding2023ApJ...955L..24G}
{Goulding}, A.~D., {Greene}, J.~E., {Setton}, D.~J., {et~al.} 2023, \bibinfo{title}{{UNCOVER: The Growth of the First Massive Black Holes from JWST/NIRSpec-Spectroscopic Redshift Confirmation of an X-Ray Luminous AGN at z = 10.1},} \apjl, 955, L24, \dodoi{10.3847/2041-8213/acf7c5}

\bibitem[{J.~E. {Greene} {et~al.}(2024){Greene}, {Labbe}, {Goulding}, {Furtak}, {Chemerynska}, {Kokorev}, {Dayal}, {Volonteri}, {Williams}, {Wang}, {Setton}, {Burgasser}, {Bezanson}, {Atek}, {Brammer}, {Cutler}, {Feldmann}, {Fujimoto}, {Glazebrook}, {de Graaff}, {Khullar}, {Leja}, {Marchesini}, {Maseda}, {Matthee}, {Miller}, {Naidu}, {Nanayakkara}, {Oesch}, {Pan}, {Papovich}, {Price}, {van Dokkum}, {Weaver}, {Whitaker}, \& {Zitrin}}]{Greene2024}
{Greene}, J.~E., {Labbe}, I., {Goulding}, A.~D., {et~al.} 2024, \bibinfo{title}{{UNCOVER Spectroscopy Confirms the Surprising Ubiquity of Active Galactic Nuclei in Red Sources at z > 5},} \apj, 964, 39, \dodoi{10.3847/1538-4357/ad1e5f}

\bibitem[{L. {Haemmerl{\'e}}(2021){Haemmerl{\'e}}}]{Haemmerle2021}
{Haemmerl{\'e}}, L. 2021, \bibinfo{title}{{Establishing a reliable determination of the final mass for rapidly accreting supermassive stars},} \aap, 647, A83, \dodoi{10.1051/0004-6361/202039686}

\bibitem[{L. {Haemmerl{\'e}} {et~al.}(2018){Haemmerl{\'e}}, {Woods}, {Klessen}, {Heger}, \& {Whalen}}]{Haemmerle2018}
{Haemmerl{\'e}}, L., {Woods}, T.~E., {Klessen}, R.~S., {Heger}, A., \& {Whalen}, D.~J. 2018, \bibinfo{title}{{The evolution of supermassive Population III stars},} \mnras, 474, 2757, \dodoi{10.1093/mnras/stx2919}

\bibitem[{O. {Hahn} \& T. {Abel}(2011){Hahn} \& {Abel}}]{hahn2011multi}
{Hahn}, O., \& {Abel}, T. 2011, \bibinfo{title}{{Multi-scale initial conditions for cosmological simulations},} \mnras, 415, 2101, \dodoi{10.1111/j.1365-2966.2011.18820.x}

\bibitem[{S. Hawking(1971)Hawking}]{hawking1971gravitationally}
Hawking, S. 1971, \bibinfo{title}{Gravitationally collapsed objects of very low mass,} Monthly Notices of the Royal Astronomical Society, 152, 75

\bibitem[{N.~P. {Herrington} {et~al.}(2023){Herrington}, {Whalen}, \& {Woods}}]{Herrington2023}
{Herrington}, N.~P., {Whalen}, D.~J., \& {Woods}, T.~E. 2023, \bibinfo{title}{{Modelling supermassive primordial stars with MESA},} \mnras, 521, 463, \dodoi{10.1093/mnras/stad572}

\bibitem[{S. {Hirano} {et~al.}(2018){Hirano}, {Yoshida}, {Sakurai}, \& {Fujii}}]{Hirano2018ApJ...855...17H}
{Hirano}, S., {Yoshida}, N., {Sakurai}, Y., \& {Fujii}, M.~S. 2018, \bibinfo{title}{{Formation of the First Star Clusters and Massive Star Binaries by Fragmentation of Filamentary Primordial Gas Clouds},} \apj, 855, 17, \dodoi{10.3847/1538-4357/aaaaba}

\bibitem[{S. {Hirano} {et~al.}(2015){Hirano}, {Zhu}, {Yoshida}, {Spergel}, \& {Yorke}}]{Hirano2015ApJ...814...18H}
{Hirano}, S., {Zhu}, N., {Yoshida}, N., {Spergel}, D., \& {Yorke}, H.~W. 2015, \bibinfo{title}{{Early Structure Formation from Primordial Density Fluctuations with a Blue, Tilted Power Spectrum},} \apj, 814, 18, \dodoi{10.1088/0004-637X/814/1/18}

\bibitem[{G. {Hobbs} \& S. {Dai}(2017){Hobbs} \& {Dai}}]{Hobbs2017NSRev...4..707H}
{Hobbs}, G., \& {Dai}, S. 2017, \bibinfo{title}{{Gravitational wave research using pulsar timing arrays},} National Science Review, 4, 707, \dodoi{10.1093/nsr/nwx126}

\bibitem[{D. {Hooper} {et~al.}(2024){Hooper}, {Ireland}, {Krnjaic}, \& {Stebbins}}]{Hooper2024}
{Hooper}, D., {Ireland}, A., {Krnjaic}, G., \& {Stebbins}, A. 2024, \bibinfo{title}{{Supermassive primordial black holes from inflation},} \jcap, 2024, 021, \dodoi{10.1088/1475-7516/2024/04/021}

\bibitem[{P.~F. {Hopkins}(2015){Hopkins}}]{Hopkins2015MNRAS.450...53H}
{Hopkins}, P.~F. 2015, \bibinfo{title}{{A new class of accurate, mesh-free hydrodynamic simulation methods},} \mnras, 450, 53, \dodoi{10.1093/mnras/stv195}

\bibitem[{T. {Hosokawa} {et~al.}(2013){Hosokawa}, {Yorke}, {Inayoshi}, {Omukai}, \& {Yoshida}}]{Hosokawa2013}
{Hosokawa}, T., {Yorke}, H.~W., {Inayoshi}, K., {Omukai}, K., \& {Yoshida}, N. 2013, \bibinfo{title}{{Formation of Primordial Supermassive Stars by Rapid Mass Accretion},} \apj, 778, 178, \dodoi{10.1088/0004-637X/778/2/178}

\bibitem[{H.-L. {Huang} {et~al.}(2024){Huang}, {Jiang}, \& {Piao}}]{Huang2024PhRvD.110j3540H}
{Huang}, H.-L., {Jiang}, J.-Q., \& {Piao}, Y.-S. 2024, \bibinfo{title}{{High-redshift JWST massive galaxies and the initial clustering of supermassive primordial black holes},} \prd, 110, 103540, \dodoi{10.1103/PhysRevD.110.103540}

\bibitem[{K. {Inayoshi} {et~al.}(2018){Inayoshi}, {Ichikawa}, \& {Haiman}}]{InayoshiSMBHGW2018ApJ}
{Inayoshi}, K., {Ichikawa}, K., \& {Haiman}, Z. 2018, \bibinfo{title}{{Gravitational Waves from Supermassive Black Hole Binaries in Ultraluminous Infrared Galaxies},} \apjl, 863, L36, \dodoi{10.3847/2041-8213/aad8ad}

\bibitem[{K. {Inayoshi} {et~al.}(2024){Inayoshi}, {Kashiyama}, {Li}, {Harikane}, {Ichikawa}, \& {Onoue}}]{Inayoshi2024ApJ...966..164I}
{Inayoshi}, K., {Kashiyama}, K., {Li}, W., {et~al.} 2024, \bibinfo{title}{{Exploring Low-mass Black Holes through Tidal Disruption Events in the Early Universe: Perspectives in the Era of the JWST, Roman Space Telescope, and LSST Surveys},} \apj, 966, 164, \dodoi{10.3847/1538-4357/ad344c}

\bibitem[{K. {Inayoshi} {et~al.}(2014){Inayoshi}, {Omukai}, \& {Tasker}}]{InayoshiSMS2014MNRAS}
{Inayoshi}, K., {Omukai}, K., \& {Tasker}, E. 2014, \bibinfo{title}{{Formation of an embryonic supermassive star in the first galaxy.},} \mnras, 445, L109, \dodoi{10.1093/mnrasl/slu151}

\bibitem[{K. {Inayoshi} {et~al.}(2025){Inayoshi}, {Shangguan}, {Chen}, {Ho}, \& {Haiman}}]{InayoshiBinaryBH2025arXiv}
{Inayoshi}, K., {Shangguan}, J., {Chen}, X., {Ho}, L.~C., \& {Haiman}, Z. 2025, \bibinfo{title}{{The Emergence of Little Red Dots from Binary Massive Black Holes},} arXiv e-prints, arXiv:2505.05322, \dodoi{10.48550/arXiv.2505.05322}

\bibitem[{K. {Inayoshi} {et~al.}(2020){Inayoshi}, {Visbal}, \& {Haiman}}]{Inayoshi:2020}
{Inayoshi}, K., {Visbal}, E., \& {Haiman}, Z. 2020, \bibinfo{title}{{The Assembly of the First Massive Black Holes},} \araa, 58, 27, \dodoi{10.1146/annurev-astro-120419-014455}

\bibitem[{D. {Inman} \& Y. {Ali-Ha{\"\i}moud}(2019){Inman} \& {Ali-Ha{\"\i}moud}}]{Inman2019PhRvD.100h3528I}
{Inman}, D., \& {Ali-Ha{\"\i}moud}, Y. 2019, \bibinfo{title}{{Early structure formation in primordial black hole cosmologies},} \prd, 100, 083528, \dodoi{10.1103/PhysRevD.100.083528}

\bibitem[{M. {Ito} \& K. {Omukai}(2024){Ito} \& {Omukai}}]{Ito2024PASJ...76..850I}
{Ito}, M., \& {Omukai}, K. 2024, \bibinfo{title}{{First star formation in extremely early epochs},} \pasj, 76, 850, \dodoi{10.1093/pasj/psae054}

\bibitem[{H. {Jiao} {et~al.}(2024){Jiao}, {Brandenberger}, \& {Refregier}}]{Jiao2024}
{Jiao}, H., {Brandenberger}, R., \& {Refregier}, A. 2024, \bibinfo{title}{{N -body simulation of early structure formation from cosmic string loops},} \prd, 109, 123524, \dodoi{10.1103/PhysRevD.109.123524}

\bibitem[{J.~L. {Johnson} \& V. {Bromm}(2006){Johnson} \& {Bromm}}]{Johnson2006MNRAS.366..247J}
{Johnson}, J.~L., \& {Bromm}, V. 2006, \bibinfo{title}{{The cooling of shock-compressed primordial gas},} \mnras, 366, 247, \dodoi{10.1111/j.1365-2966.2005.09846.x}

\bibitem[{A. {Kashlinsky}(2021){Kashlinsky}}]{Kashlinsky2021PhRvL.126a1101K}
{Kashlinsky}, A. 2021, \bibinfo{title}{{Cosmological Advection Flows in the Presence of Primordial Black Holes as Dark Matter and Formation of First Sources},} \prl, 126, 011101, \dodoi{10.1103/PhysRevLett.126.011101}

\bibitem[{M. {Kawasaki} \& K. {Murai}(2019){Kawasaki} \& {Murai}}]{Kawasaki2019PhRvD.100j3521K}
{Kawasaki}, M., \& {Murai}, K. 2019, \bibinfo{title}{{Formation of supermassive primordial black holes by Affleck-Dine mechanism},} \prd, 100, 103521, \dodoi{10.1103/PhysRevD.100.103521}

\bibitem[{R.~S. {Klessen} \& S.~C.~O. {Glover}(2023){Klessen} \& {Glover}}]{Klessen:2023FirstStars}
{Klessen}, R.~S., \& {Glover}, S. C.~O. 2023, \bibinfo{title}{{The First Stars: Formation, Properties, and Impact},} \araa, 61, 65, \dodoi{10.1146/annurev-astro-071221-053453}

\bibitem[{D.~D. {Kocevski} {et~al.}(2025){Kocevski}, {Finkelstein}, {Barro}, {Taylor}, {Calabr{\`o}}, {Laloux}, {Buchner}, {Trump}, {Leung}, {Yang}, {Dickinson}, {P{\'e}rez-Gonz{\'a}lez}, {Pacucci}, {Inayoshi}, {Somerville}, {McGrath}, {Akins}, {Bagley}, {Bowler}, {Bisigello}, {Carnall}, {Casey}, {Cheng}, {Cleri}, {Costantin}, {Cullen}, {Davis}, {Donnan}, {Dunlop}, {Ellis}, {Ferguson}, {Fujimoto}, {Fontana}, {Giavalisco}, {Grazian}, {Grogin}, {Hathi}, {Hirschmann}, {Huertas-Company}, {Holwerda}, {Illingworth}, {Juneau}, {Kartaltepe}, {Koekemoer}, {Li}, {Lucas}, {Magee}, {Mason}, {McLeod}, {McLure}, {Napolitano}, {Papovich}, {Pirzkal}, {Rodighiero}, {Santini}, {Wilkins}, \& {Yung}}]{KocevskiLRD2025ApJ...986..126K}
{Kocevski}, D.~D., {Finkelstein}, S.~L., {Barro}, G., {et~al.} 2025, \bibinfo{title}{{The Rise of Faint, Red Active Galactic Nuclei at z > 4: A Sample of Little Red Dots in the JWST Extragalactic Legacy Fields},} \apj, 986, 126, \dodoi{10.3847/1538-4357/adbc7d}

\bibitem[{A. {Kogut} {et~al.}(2025){Kogut}, {Aghanim}, {Chluba}, {Chuss}, {Delabrouille}, {Dvorkin}, {Fixsen}, {Ghosh}, {Hensley}, {Hill}, {Maffei}, {Pullen}, {Rotti}, {Sabyr}, {Switzer}, {Thiele}, {Wollack}, \& {Zelko}}]{PIXIE2025JCAP...04..020K}
{Kogut}, A., {Aghanim}, N., {Chluba}, J., {et~al.} 2025, \bibinfo{title}{{The Primordial Inflation Explorer (PIXIE): mission design and science goals},} \jcap, 2025, 020, \dodoi{10.1088/1475-7516/2025/04/020}

\bibitem[{K. {Kohri} {et~al.}(2022){Kohri}, {Sekiguchi}, \& {Wang}}]{Kohri2022PhRvD.106d3539K}
{Kohri}, K., {Sekiguchi}, T., \& {Wang}, S. 2022, \bibinfo{title}{{Cosmological 21-cm line observations to test scenarios of super-Eddington accretion on to black holes being seeds of high-redshifted supermassive black holes},} \prd, 106, 043539, \dodoi{10.1103/PhysRevD.106.043539}

\bibitem[{V. {Kokorev} {et~al.}(2024){Kokorev}, {Caputi}, {Greene}, {Dayal}, {Trebitsch}, {Cutler}, {Fujimoto}, {Labb{\'e}}, {Miller}, {Iani}, {Navarro-Carrera}, \& {Rinaldi}}]{KokorvLRD2024ApJ...968...38K}
{Kokorev}, V., {Caputi}, K.~I., {Greene}, J.~E., {et~al.} 2024, \bibinfo{title}{{A Census of Photometrically Selected Little Red Dots at 4 < z < 9 in JWST Blank Fields},} \apj, 968, 38, \dodoi{10.3847/1538-4357/ad4265}

\bibitem[{O.~E. {Kov{\'a}cs} {et~al.}(2024){Kov{\'a}cs}, {Bogd{\'a}n}, {Natarajan}, {Werner}, {Azadi}, {Volonteri}, {Tremblay}, {Chadayammuri}, {Forman}, {Jones}, \& {Kraft}}]{Kovacs2024ApJ...965L..21K}
{Kov{\'a}cs}, O.~E., {Bogd{\'a}n}, {\'A}., {Natarajan}, P., {et~al.} 2024, \bibinfo{title}{{A Candidate Supermassive Black Hole in a Gravitationally Lensed Galaxy at Z {\ensuremath{\approx}} 10},} \apjl, 965, L21, \dodoi{10.3847/2041-8213/ad391f}

\bibitem[{I. {Labb{\'e}} {et~al.}(2023){Labb{\'e}}, {van Dokkum}, {Nelson}, {Bezanson}, {Suess}, {Leja}, {Brammer}, {Whitaker}, {Mathews}, {Stefanon}, \& {Wang}}]{Labbe2023Natur.616..266L}
{Labb{\'e}}, I., {van Dokkum}, P., {Nelson}, E., {et~al.} 2023, \bibinfo{title}{{A population of red candidate massive galaxies 600 Myr after the Big Bang},} \nat, 616, 266, \dodoi{10.1038/s41586-023-05786-2}

\bibitem[{I. {Labbe} {et~al.}(2025){Labbe}, {Greene}, {Bezanson}, {Fujimoto}, {Furtak}, {Goulding}, {Matthee}, {Naidu}, {Oesch}, {Atek}, {Brammer}, {Chemerynska}, {Coe}, {Cutler}, {Dayal}, {Feldmann}, {Franx}, {Glazebrook}, {Leja}, {Maseda}, {Marchesini}, {Nanayakkara}, {Nelson}, {Pan}, {Papovich}, {Price}, {Suess}, {Wang}, {Weaver}, {Whitaker}, {Williams}, \& {Zitrin}}]{LabbeLRD2025ApJ...978...92L}
{Labbe}, I., {Greene}, J.~E., {Bezanson}, R., {et~al.} 2025, \bibinfo{title}{{UNCOVER: Candidate Red Active Galactic Nuclei at 3 < z < 7 with JWST and ALMA},} \apj, 978, 92, \dodoi{10.3847/1538-4357/ad3551}

\bibitem[{R.~L. {Larson} {et~al.}(2023){Larson}, {Finkelstein}, {Kocevski}, {Hutchison}, {Trump}, {Arrabal Haro}, {Bromm}, {Cleri}, {Dickinson}, {Fujimoto}, {Kartaltepe}, {Koekemoer}, {Papovich}, {Pirzkal}, {Tacchella}, {Zavala}, {Bagley}, {Behroozi}, {Champagne}, {Cole}, {Jung}, {Morales}, {Yang}, {Zhang}, {Zitrin}, {Amor{\'\i}n}, {Burgarella}, {Casey}, {Ch{\'a}vez Ortiz}, {Cox}, {Chworowsky}, {Fontana}, {Gawiser}, {Grazian}, {Grogin}, {Harish}, {Hathi}, {Hirschmann}, {Holwerda}, {Juneau}, {Leung}, {Lucas}, {McGrath}, {P{\'e}rez-Gonz{\'a}lez}, {Rigby}, {Seill{\'e}}, {Simons}, {de La Vega}, {Weiner}, {Wilkins}, {Yung}, \& {Ceers Team}}]{Larson_2023_BH}
{Larson}, R.~L., {Finkelstein}, S.~L., {Kocevski}, D.~D., {et~al.} 2023, \bibinfo{title}{{A CEERS Discovery of an Accreting Supermassive Black Hole 570 Myr after the Big Bang: Identifying a Progenitor of Massive z > 6 Quasars},} \apjl, 953, L29, \dodoi{10.3847/2041-8213/ace619}

\bibitem[{M.~A. {Latif} {et~al.}(2021){Latif}, {Khochfar}, {Schleicher}, \& {Whalen}}]{Latif2021}
{Latif}, M.~A., {Khochfar}, S., {Schleicher}, D., \& {Whalen}, D.~J. 2021, \bibinfo{title}{{Radiation hydrodynamical simulations of the birth of intermediate-mass black holes in the first galaxies},} \mnras, 508, 1756, \dodoi{10.1093/mnras/stab2708}

\bibitem[{M.~A. {Latif} {et~al.}(2020){Latif}, {Khochfar}, \& {Whalen}}]{Latif2020ApJSMBHB}
{Latif}, M.~A., {Khochfar}, S., \& {Whalen}, D. 2020, \bibinfo{title}{{The Birth of Binary Direct-collapse Black Holes},} \apjl, 892, L4, \dodoi{10.3847/2041-8213/ab7c61}

\bibitem[{M.~A. {Latif} {et~al.}(2022){Latif}, {Whalen}, {Khochfar}, {Herrington}, \& {Woods}}]{Latif2022_DCBH}
{Latif}, M.~A., {Whalen}, D.~J., {Khochfar}, S., {Herrington}, N.~P., \& {Woods}, T.~E. 2022, \bibinfo{title}{{Turbulent cold flows gave birth to the first quasars},} \nat, 607, 48, \dodoi{10.1038/s41586-022-04813-y}

\bibitem[{G.~C.~K. {Leung} {et~al.}(2024){Leung}, {Finkelstein}, {P{\'e}rez-Gonz{\'a}lez}, {Morales}, {Taylor}, {Barro}, {Kocevski}, {Akins}, {Carnall}, {Ch{\'a}vez Ortiz}, {Cleri}, {Cullen}, {Donnan}, {Dunlop}, {Ellis}, {Grogin}, {Hirschmann}, {Koekemoer}, {Kokorev}, {Lucas}, {McLeod}, {Papovich}, \& {Yung}}]{Leung2024:LRDarXiv}
{Leung}, G. C.~K., {Finkelstein}, S.~L., {P{\'e}rez-Gonz{\'a}lez}, P.~G., {et~al.} 2024, \bibinfo{title}{{Exploring the Nature of Little Red Dots: Constraints on AGN and Stellar Contributions from PRIMER MIRI Imaging},} arXiv e-prints, arXiv:2411.12005.
\newblock \doarXiv{2411.12005}

\bibitem[{E.-K. {Li} {et~al.}(2025){Li}, {Liu}, {Torres-Orjuela}, {Chen}, {Inayoshi}, {Wang}, {Hu}, {Amaro-Seoane}, {Askar}, {Bambi}, {Capelo}, {Chen}, {Chua}, {Cond{\'e}s-Bre{\~n}a}, {Dai}, {Das}, {Derdzinski}, {Fan}, {Fujii}, {Gao}, {Garg}, {Ge}, {Giersz}, {Huang}, {Hypki}, {Liang}, {Liu}, {Liu}, {Liu}, {Liu}, {Mayer}, {Napolitano}, {Peng}, {Shao}, {Shashank}, {Shen}, {Tagawa}, {Tanikawa}, {Toscani}, {V{\'a}zquez-Aceves}, {Wang}, {Wang}, {Yi}, {Zhang}, {Zhang}, {Zhu}, {Zwick}, {Huang}, {Mei}, {Wang}, {Xie}, {Zhang}, \& {Luo}}]{TianQin2025RPPh...88e6901L}
{Li}, E.-K., {Liu}, S., {Torres-Orjuela}, A., {et~al.} 2025, \bibinfo{title}{{Gravitational wave astronomy with TianQin},} Reports on Progress in Physics, 88, 056901, \dodoi{10.1088/1361-6633/adc9be}

\bibitem[{B. {Liu} \& V. {Bromm}(2018){Liu} \& {Bromm}}]{LiuBromm2018}
{Liu}, B., \& {Bromm}, V. 2018, \bibinfo{title}{{Effect of lithium hydride on the cooling of primordial gas},} \mnras, 476, 1826, \dodoi{10.1093/mnras/sty350}

\bibitem[{B. {Liu} \& V. {Bromm}(2022){Liu} \& {Bromm}}]{Boyuan2022ApJ}
{Liu}, B., \& {Bromm}, V. 2022, \bibinfo{title}{{Accelerating Early Massive Galaxy Formation with Primordial Black Holes},} \apjl, 937, L30, \dodoi{10.3847/2041-8213/ac927f}

\bibitem[{B. {Liu} \& V. {Bromm}(2023){Liu} \& {Bromm}}]{Boyuan2023arXiv231204085L}
{Liu}, B., \& {Bromm}, V. 2023, \bibinfo{title}{{Impact of primordial black holes on the formation of the first stars and galaxies},} arXiv e-prints, arXiv:2312.04085, \dodoi{10.48550/arXiv.2312.04085}

\bibitem[{B. {Liu} {et~al.}(2024){Liu}, {Gurian}, {Inayoshi}, {Hirano}, {Hosokawa}, {Bromm}, \& {Yoshida}}]{Liu2024_Mass}
{Liu}, B., {Gurian}, J., {Inayoshi}, K., {et~al.} 2024, \bibinfo{title}{{Towards a universal analytical model for Population III star formation: interplay between feedback and fragmentation},} \mnras, 534, 290, \dodoi{10.1093/mnras/stae2066}

\bibitem[{B. {Liu} {et~al.}(2022){Liu}, {Zhang}, \& {Bromm}}]{Boyuan2022MNRAS.514.2376L}
{Liu}, B., {Zhang}, S., \& {Bromm}, V. 2022, \bibinfo{title}{{Effects of stellar-mass primordial black holes on first star formation},} \mnras, 514, 2376, \dodoi{10.1093/mnras/stac1472}

\bibitem[{S. {Liu} {et~al.}(2024){Liu}, {Wang}, {Hu}, {Tanikawa}, \& {Trani}}]{Liu2024sc}
{Liu}, S., {Wang}, L., {Hu}, Y.-M., {Tanikawa}, A., \& {Trani}, A.~A. 2024, \bibinfo{title}{{Merging hierarchical triple black hole systems with intermediate-mass black holes in population III star clusters},} \mnras, 533, 2262, \dodoi{10.1093/mnras/stae1946}

\bibitem[{G. {Lodato} \& P. {Natarajan}(2006){Lodato} \& {Natarajan}}]{Lodato:2006DCBH}
{Lodato}, G., \& {Natarajan}, P. 2006, \bibinfo{title}{{Supermassive black hole formation during the assembly of pre-galactic discs},} \mnras, 371, 1813, \dodoi{10.1111/j.1365-2966.2006.10801.x}

\bibitem[{A. {Loeb} \& F.~A. {Rasio}(1994){Loeb} \& {Rasio}}]{Loeb1994ApJ...432...52L}
{Loeb}, A., \& {Rasio}, F.~A. 1994, \bibinfo{title}{{Collapse of Primordial Gas Clouds and the Formation of Quasar Black Holes},} \apj, 432, 52, \dodoi{10.1086/174548}

\bibitem[{P. {Lu} {et~al.}(2021){Lu}, {Takhistov}, {Gelmini}, {Hayashi}, {Inoue}, \& {Kusenko}}]{Lu2021}
{Lu}, P., {Takhistov}, V., {Gelmini}, G.~B., {et~al.} 2021, \bibinfo{title}{{Constraining Primordial Black Holes with Dwarf Galaxy Heating},} \apjl, 908, L23, \dodoi{10.3847/2041-8213/abdcb6}

\bibitem[{Y. {Lu} {et~al.}(2024){Lu}, {Picker}, \& {Kusenko}}]{Lu2024PhRvD.109l3016L}
{Lu}, Y., {Picker}, Z. S.~C., \& {Kusenko}, A. 2024, \bibinfo{title}{{High-redshift supermassive black holes from tiny black hole explosions},} \prd, 109, 123016, \dodoi{10.1103/PhysRevD.109.123016}

\bibitem[{J. {Luo} {et~al.}(2016){Luo}, {Chen}, {Duan}, {Gong}, {Hu}, {Ji}, {Liu}, {Mei}, {Milyukov}, {Sazhin}, {Shao}, {Toth}, {Tu}, {Wang}, {Wang}, {Yeh}, {Zhan}, {Zhang}, {Zharov}, \& {Zhou}}]{TianQin2016CQGra..33c5010L}
{Luo}, J., {Chen}, L.-S., {Duan}, H.-Z., {et~al.} 2016, \bibinfo{title}{{TianQin: a space-borne gravitational wave detector},} Classical and Quantum Gravity, 33, 035010, \dodoi{10.1088/0264-9381/33/3/035010}

\bibitem[{L. {Ma} {et~al.}(2021){Ma}, {Hopkins}, {Ma}, {Angl{\'e}s-Alc{\'a}zar}, {Faucher-Gigu{\`e}re}, \& {Kelley}}]{Ma2021}
{Ma}, L., {Hopkins}, P.~F., {Ma}, X., {et~al.} 2021, \bibinfo{title}{{Seeds don't sink: even massive black hole 'seeds' cannot migrate to galaxy centres efficiently},} \mnras, 508, 1973, \dodoi{10.1093/mnras/stab2713}

\bibitem[{K.~J. {Mack} {et~al.}(2007){Mack}, {Ostriker}, \& {Ricotti}}]{Mack2007ApJ}
{Mack}, K.~J., {Ostriker}, J.~P., \& {Ricotti}, M. 2007, \bibinfo{title}{{Growth of Structure Seeded by Primordial Black Holes},} \apj, 665, 1277, \dodoi{10.1086/518998}

\bibitem[{R. {Maiolino} {et~al.}(2024){Maiolino}, {Scholtz}, {Curtis-Lake}, {Carniani}, {Baker}, {de Graaff}, {Tacchella}, {{\"U}bler}, {D'Eugenio}, {Witstok}, {Curti}, {Arribas}, {Bunker}, {Charlot}, {Chevallard}, {Eisenstein}, {Egami}, {Ji}, {Jones}, {Lyu}, {Rawle}, {Robertson}, {Rujopakarn}, {Perna}, {Sun}, {Venturi}, {Williams}, \& {Willott}}]{Maiolino2024A&A}
{Maiolino}, R., {Scholtz}, J., {Curtis-Lake}, E., {et~al.} 2024, \bibinfo{title}{{JADES: The diverse population of infant black holes at 4 < z < 11: Merging, tiny, poor, but mighty},} \aap, 691, A145, \dodoi{10.1051/0004-6361/202347640}

\bibitem[{R. {Maiolino} {et~al.}(2025){Maiolino}, {Uebler}, {D'Eugenio}, {Scholtz}, {Juodzbalis}, {Ji}, {Perna}, {Bromm}, {Dayal}, {Koudmani}, {Liu}, {Schneider}, {Sijacki}, {Valiante}, {Trinca}, {Zhang}, {Volonteri}, {Inayoshi}, {Carniani}, {Nakajima}, {Isobe}, {Witstok}, {Jones}, {Tacchella}, {Arribas}, {Bunker}, {Cataldi}, {Charlot}, {Cresci}, {Curti}, {Fabian}, {Katz}, {Kumari}, {Laporte}, {Mazzolari}, {Robertson}, {Sun}, {Rodriguez Del Pino}, \& {Venturi}}]{Maiolino2025arXiv250522567M}
{Maiolino}, R., {Uebler}, H., {D'Eugenio}, F., {et~al.} 2025, \bibinfo{title}{{A black hole in a near-pristine galaxy 700 million years after the Big Bang},} arXiv e-prints, arXiv:2505.22567, \dodoi{10.48550/arXiv.2505.22567}

\bibitem[{Y. {Matsuoka} {et~al.}(2024){Matsuoka}, {Izumi}, {Onoue}, {Strauss}, {Iwasawa}, {Kashikawa}, {Akiyama}, {Aoki}, {Arita}, {Imanishi}, {Ishimoto}, {Kawaguchi}, {Kohno}, {Lee}, {Nagao}, {Silverman}, \& {Toba}}]{Matsuoka2024ApJ...965L...4M}
{Matsuoka}, Y., {Izumi}, T., {Onoue}, M., {et~al.} 2024, \bibinfo{title}{{Discovery of Merging Twin Quasars at z = 6.05},} \apjl, 965, L4, \dodoi{10.3847/2041-8213/ad35c7}

\bibitem[{A. {Matteri} {et~al.}(2025){Matteri}, {Ferrara}, \& {Pallottini}}]{Matteri2025arXiv250318850M}
{Matteri}, A., {Ferrara}, A., \& {Pallottini}, A. 2025, \bibinfo{title}{{Beyond the first galaxies primordial black holes shine},} arXiv e-prints, arXiv:2503.18850, \dodoi{10.48550/arXiv.2503.18850}

\bibitem[{J. {Matthee} {et~al.}(2024){Matthee}, {Naidu}, {Brammer}, {Chisholm}, {Eilers}, {Goulding}, {Greene}, {Kashino}, {Labbe}, {Lilly}, {Mackenzie}, {Oesch}, {Weibel}, {Wuyts}, {Xiao}, {Bordoloi}, {Bouwens}, {van Dokkum}, {Illingworth}, {Kramarenko}, {Maseda}, {Mason}, {Meyer}, {Nelson}, {Reddy}, {Shivaei}, {Simcoe}, \& {Yue}}]{MattheeLRD2024ApJ...963..129M}
{Matthee}, J., {Naidu}, R.~P., {Brammer}, G., {et~al.} 2024, \bibinfo{title}{{Little Red Dots: An Abundant Population of Faint Active Galactic Nuclei at z {\ensuremath{\sim}} 5 Revealed by the EIGER and FRESCO JWST Surveys},} \apj, 963, 129, \dodoi{10.3847/1538-4357/ad2345}

\bibitem[{P. {Meszaros}(1975){Meszaros}}]{Meszaros1975A&A....38....5M}
{Meszaros}, P. 1975, \bibinfo{title}{{Primeval black holes and galaxy formation.},} \aap, 38, 5

\bibitem[{M. {Milosavljevi{\'c}} \& D. {Merritt}(2003{\natexlab{a}}){Milosavljevi{\'c}} \& {Merritt}}]{MilosavljevicFinalpc2003AIPC}
{Milosavljevi{\'c}}, M., \& {Merritt}, D. 2003{\natexlab{a}}, in American Institute of Physics Conference Series, Vol. 686, The Astrophysics of Gravitational Wave Sources, ed. J.~M. {Centrella} (AIP), 201--210, \dodoi{10.1063/1.1629432}

\bibitem[{M. {Milosavljevi{\'c}} \& D. {Merritt}(2003{\natexlab{b}}){Milosavljevi{\'c}} \& {Merritt}}]{MilosavljevicMBHB2003ApJ}
{Milosavljevi{\'c}}, M., \& {Merritt}, D. 2003{\natexlab{b}}, \bibinfo{title}{{Long-Term Evolution of Massive Black Hole Binaries},} \apj, 596, 860, \dodoi{10.1086/378086}

\bibitem[{M. {Mirbabayi} {et~al.}(2020){Mirbabayi}, {Gruzinov}, \& {Nore{\~n}a}}]{Mirbabayi2020}
{Mirbabayi}, M., {Gruzinov}, A., \& {Nore{\~n}a}, J. 2020, \bibinfo{title}{{Spin of primordial black holes},} \jcap, 2020, 017, \dodoi{10.1088/1475-7516/2020/03/017}

\bibitem[{C. {Nagele} {et~al.}(2022){Nagele}, {Umeda}, {Takahashi}, {Yoshida}, \& {Sumiyoshi}}]{Nagele2022}
{Nagele}, C., {Umeda}, H., {Takahashi}, K., {Yoshida}, T., \& {Sumiyoshi}, K. 2022, \bibinfo{title}{{Stability analysis of supermassive primordial stars: a new mass range for general relativistic instability supernovae},} \mnras, 517, 1584, \dodoi{10.1093/mnras/stac2495}

\bibitem[{T. {Nakama} {et~al.}(2018){Nakama}, {Carr}, \& {Silk}}]{Nakama2018PhRvD..97d3525N}
{Nakama}, T., {Carr}, B., \& {Silk}, J. 2018, \bibinfo{title}{{Limits on primordial black holes from {\ensuremath{\mu}} distortions in cosmic microwave background},} \prd, 97, 043525, \dodoi{10.1103/PhysRevD.97.043525}

\bibitem[{S. {Naoz} \& Z. {Haiman}(2023){Naoz} \& {Haiman}}]{Naoz2023ApJ...955L..27N}
{Naoz}, S., \& {Haiman}, Z. 2023, \bibinfo{title}{{The Enhanced Population of Extreme Mass-ratio Inspirals in the LISA Band from Supermassive Black Hole Binaries},} \apjl, 955, L27, \dodoi{10.3847/2041-8213/acf8c9}

\bibitem[{L. {Napolitano} {et~al.}(2025){Napolitano}, {Castellano}, {Pentericci}, {Vignali}, {Gilli}, {Fontana}, {Santini}, {Treu}, {Calabr{\`o}}, {Llerena}, {Piconcelli}, {Zappacosta}, {Mascia}, {Tripodi}, {Arrabal Haro}, {Bergamini}, {Bakx}, {Dickinson}, {Glazebrook}, {Henry}, {Leethochawalit}, {Mazzolari}, {Merlin}, {Morishita}, {Nanayakkara}, {Paris}, {Puccetti}, {Roberts-Borsani}, {Rojas Ruiz}, {Rosati}, {Vanzella}, {Vito}, {Vulcani}, {Wang}, {Yoon}, \& {Zavala}}]{GHZ9Napolitano2025ApJ...989...75N}
{Napolitano}, L., {Castellano}, M., {Pentericci}, L., {et~al.} 2025, \bibinfo{title}{{The Dual Nature of GHZ9: Coexisting Active Galactic Nuclei and Star Formation Activity in a Remote X-Ray Source at z = 10.145},} \apj, 989, 75, \dodoi{10.3847/1538-4357/ade706}

\bibitem[{P. {Natarajan} {et~al.}(2024){Natarajan}, {Pacucci}, {Ricarte}, {Bogd{\'a}n}, {Goulding}, \& {Cappelluti}}]{Natarajan:2023UHZ1}
{Natarajan}, P., {Pacucci}, F., {Ricarte}, A., {et~al.} 2024, \bibinfo{title}{{First Detection of an Overmassive Black Hole Galaxy UHZ1: Evidence for Heavy Black Hole Seed Formation from Direct Collapse},} \apjl, 960, L1, \dodoi{10.3847/2041-8213/ad0e76}

\bibitem[{A. {Negri} \& M. {Volonteri}(2017){Negri} \& {Volonteri}}]{Negri2017MNRAS.467.3475N}
{Negri}, A., \& {Volonteri}, M. 2017, \bibinfo{title}{{Black hole feeding and feedback: the physics inside the `sub-grid'},} \mnras, 467, 3475, \dodoi{10.1093/mnras/stx362}

\bibitem[{S.~P. {Oh} \& Z. {Haiman}(2002){Oh} \& {Haiman}}]{OhHaiman2002}
{Oh}, S.~P., \& {Haiman}, Z. 2002, \bibinfo{title}{{Second-Generation Objects in the Universe: Radiative Cooling and Collapse of Halos with Virial Temperatures above {}10$^{4}$ K},} \apj, 569, 558, \dodoi{10.1086/339393}

\bibitem[{ {Planck Collaboration} {et~al.}(2020){Planck Collaboration}, {Aghanim}, {Akrami}, {Ashdown}, {Aumont}, {Baccigalupi}, {Ballardini}, {Banday}, {Barreiro}, {Bartolo}, {Basak}, {Battye}, {Benabed}, {Bernard}, {Bersanelli}, {Bielewicz}, {Bock}, {Bond}, {Borrill}, {Bouchet}, {Boulanger}, {Bucher}, {Burigana}, {Butler}, {Calabrese}, {Cardoso}, {Carron}, {Challinor}, {Chiang}, {Chluba}, {Colombo}, {Combet}, {Contreras}, {Crill}, {Cuttaia}, {de Bernardis}, {de Zotti}, {Delabrouille}, {Delouis}, {Di Valentino}, {Diego}, {Dor{\'e}}, {Douspis}, {Ducout}, {Dupac}, {Dusini}, {Efstathiou}, {Elsner}, {En{\ss}lin}, {Eriksen}, {Fantaye}, {Farhang}, {Fergusson}, {Fernandez-Cobos}, {Finelli}, {Forastieri}, {Frailis}, {Fraisse}, {Franceschi}, {Frolov}, {Galeotta}, {Galli}, {Ganga}, {G{\'e}nova-Santos}, {Gerbino}, {Ghosh}, {Gonz{\'a}lez-Nuevo}, {G{\'o}rski}, {Gratton}, {Gruppuso}, {Gudmundsson}, {Hamann}, {Handley}, {Hansen}, {Herranz}, {Hildebrandt}, {Hivon}, {Huang}, {Jaffe}, {Jones}, {Karakci}, {Keih{\"a}nen},
  {Keskitalo}, {Kiiveri}, {Kim}, {Kisner}, {Knox}, {Krachmalnicoff}, {Kunz}, {Kurki-Suonio}, {Lagache}, {Lamarre}, {Lasenby}, {Lattanzi}, {Lawrence}, {Le Jeune}, {Lemos}, {Lesgourgues}, {Levrier}, {Lewis}, {Liguori}, {Lilje}, {Lilley}, {Lindholm}, {L{\'o}pez-Caniego}, {Lubin}, {Ma}, {Mac{\'\i}as-P{\'e}rez}, {Maggio}, {Maino}, {Mandolesi}, {Mangilli}, {Marcos-Caballero}, {Maris}, {Martin}, {Martinelli}, {Mart{\'\i}nez-Gonz{\'a}lez}, {Matarrese}, {Mauri}, {McEwen}, {Meinhold}, {Melchiorri}, {Mennella}, {Migliaccio}, {Millea}, {Mitra}, {Miville-Desch{\^e}nes}, {Molinari}, {Montier}, {Morgante}, {Moss}, {Natoli}, {N{\o}rgaard-Nielsen}, {Pagano}, {Paoletti}, {Partridge}, {Patanchon}, {Peiris}, {Perrotta}, {Pettorino}, {Piacentini}, {Polastri}, {Polenta}, {Puget}, {Rachen}, {Reinecke}, {Remazeilles}, {Renzi}, {Rocha}, {Rosset}, {Roudier}, {Rubi{\~n}o-Mart{\'\i}n}, {Ruiz-Granados}, {Salvati}, {Sandri}, {Savelainen}, {Scott}, {Shellard}, {Sirignano}, {Sirri}, {Spencer}, {Sunyaev}, {Suur-Uski}, {Tauber}, {Tavagnacco},
  {Tenti}, {Toffolatti}, {Tomasi}, {Trombetti}, {Valenziano}, {Valiviita}, {Van Tent}, {Vibert}, {Vielva}, {Villa}, {Vittorio}, {Wandelt}, {Wehus}, {White}, {White}, {Zacchei}, \& {Zonca}}]{Plank2020A&A...641A...6P}
{Planck Collaboration}, {Aghanim}, N., {Akrami}, Y., {et~al.} 2020, \bibinfo{title}{{Planck 2018 results. VI. Cosmological parameters},} \aap, 641, A6, \dodoi{10.1051/0004-6361/201833910}

\bibitem[{L.~{\v{C}}. {Popovi{\'c}}(2012){Popovi{\'c}}}]{PopovicSMBBH2012NewAR..56...74P}
{Popovi{\'c}}, L.~{\v{C}}. 2012, \bibinfo{title}{{Super-massive binary black holes and emission lines in active galactic nuclei},} \nar, 56, 74, \dodoi{10.1016/j.newar.2011.11.001}

\bibitem[{X. Pritchard {et~al.}(2025)Pritchard, Byrnes, Lesgourgues, \& Sharma}]{Pritchard:2025yda}
Pritchard, X., Byrnes, C.~T., Lesgourgues, J., \& Sharma, D. 2025, \bibinfo{title}{{Robust {\ensuremath{\mu}}-distortion constraints on primordial supermassive black holes from cubic (gNL) non-Gaussian perturbations},} JCAP, 07, 079, \dodoi{10.1088/1475-7516/2025/07/079}

\bibitem[{L.~R. {Prole} {et~al.}(2022){Prole}, {Clark}, {Klessen}, {Glover}, \& {Pakmor}}]{Prole2022}
{Prole}, L.~R., {Clark}, P.~C., {Klessen}, R.~S., {Glover}, S. C.~O., \& {Pakmor}, R. 2022, \bibinfo{title}{{Primordial magnetic fields in Population III star formation: a magnetized resolution study},} \mnras, 516, 2223, \dodoi{10.1093/mnras/stac2327}

\bibitem[{L.~R. {Prole} {et~al.}(2025){Prole}, {Regan}, {Mehta}, {Coles}, \& {Dayal}}]{ProlePBH2025arXiv250611233P}
{Prole}, L.~R., {Regan}, J.~A., {Mehta}, D., {Coles}, P., \& {Dayal}, P. 2025, \bibinfo{title}{{Primordial black holes in cosmological simulations: growth prospects for supermassive black holes},} arXiv e-prints, arXiv:2506.11233, \dodoi{10.48550/arXiv.2506.11233}

\bibitem[{W. {Qin} {et~al.}(2025){Qin}, {Kumar}, {Natarajan}, \& {Weiner}}]{Qin2025arXiv250613858Q}
{Qin}, W., {Kumar}, S., {Natarajan}, P., \& {Weiner}, N. 2025, \bibinfo{title}{{Not-quite-primordial black holes},} arXiv e-prints, arXiv:2506.13858, \dodoi{10.48550/arXiv.2506.13858}

\bibitem[{D.~J. {Reardon} {et~al.}(2023){Reardon}, {Zic}, {Shannon}, {Hobbs}, {Bailes}, {Di Marco}, {Kapur}, {Rogers}, {Thrane}, {Askew}, {Bhat}, {Cameron}, {Cury{\l}o}, {Coles}, {Dai}, {Goncharov}, {Kerr}, {Kulkarni}, {Levin}, {Lower}, {Manchester}, {Mandow}, {Miles}, {Nathan}, {Os{\l}owski}, {Russell}, {Spiewak}, {Zhang}, \& {Zhu}}]{PPTA2023ApJ}
{Reardon}, D.~J., {Zic}, A., {Shannon}, R.~M., {et~al.} 2023, \bibinfo{title}{{Search for an Isotropic Gravitational-wave Background with the Parkes Pulsar Timing Array},} \apjl, 951, L6, \dodoi{10.3847/2041-8213/acdd02}

\bibitem[{J. {Regan} \& M. {Volonteri}(2024){Regan} \& {Volonteri}}]{Regan2024}
{Regan}, J., \& {Volonteri}, M. 2024, \bibinfo{title}{{Massive Black Hole Seeds},} The Open Journal of Astrophysics, 7, 72, \dodoi{10.33232/001c.123239}

\bibitem[{J.~A. {Regan} {et~al.}(2020){Regan}, {Wise}, {Woods}, {Downes}, {O'Shea}, \& {Norman}}]{Regan2020}
{Regan}, J.~A., {Wise}, J.~H., {Woods}, T.~E., {et~al.} 2020, \bibinfo{title}{{The Formation of Very Massive Stars in Early Galaxies and Implications for Intermediate Mass Black Holes},} The Open Journal of Astrophysics, 3, 15, \dodoi{10.21105/astro.2008.08090}

\bibitem[{B. {Reinoso} {et~al.}(2023){Reinoso}, {Klessen}, {Schleicher}, {Glover}, \& {Solar}}]{Reinoso2023}
{Reinoso}, B., {Klessen}, R.~S., {Schleicher}, D., {Glover}, S. C.~O., \& {Solar}, P. 2023, \bibinfo{title}{{Formation of supermassive stars in the first star clusters},} \mnras, 521, 3553, \dodoi{10.1093/mnras/stad790}

\bibitem[{C. {Reisswig} {et~al.}(2013){Reisswig}, {Ott}, {Abdikamalov}, {Haas}, {M{\"o}sta}, \& {Schnetter}}]{Reisswig2013_DCBH}
{Reisswig}, C., {Ott}, C.~D., {Abdikamalov}, E., {et~al.} 2013, \bibinfo{title}{{Formation and Coalescence of Cosmological Supermassive-Black-Hole Binaries in Supermassive-Star Collapse},} \prl, 111, 151101, \dodoi{10.1103/PhysRevLett.111.151101}

\bibitem[{M. {Ricotti} {et~al.}(2008){Ricotti}, {Ostriker}, \& {Mack}}]{Ricotti2008ApJII}
{Ricotti}, M., {Ostriker}, J.~P., \& {Mack}, K.~J. 2008, \bibinfo{title}{{Effect of Primordial Black Holes on the Cosmic Microwave Background and Cosmological Parameter Estimates},} \apj, 680, 829, \dodoi{10.1086/587831}

\bibitem[{C. {Roedig} {et~al.}(2014){Roedig}, {Krolik}, \& {Miller}}]{RoedigSMBHB2014ApJ}
{Roedig}, C., {Krolik}, J.~H., \& {Miller}, M.~C. 2014, \bibinfo{title}{{Observational Signatures of Binary Supermassive Black Holes},} \apj, 785, 115, \dodoi{10.1088/0004-637X/785/2/115}

\bibitem[{J.~D. {Romano} \& N.~J. {Cornish}(2017){Romano} \& {Cornish}}]{Romano2017LRR....20....2R}
{Romano}, J.~D., \& {Cornish}, N.~J. 2017, \bibinfo{title}{{Detection methods for stochastic gravitational-wave backgrounds: a unified treatment},} Living Reviews in Relativity, 20, 2, \dodoi{10.1007/s41114-017-0004-1}

\bibitem[{Y. {Sakurai} {et~al.}(2017){Sakurai}, {Yoshida}, {Fujii}, \& {Hirano}}]{Sakurai2017}
{Sakurai}, Y., {Yoshida}, N., {Fujii}, M.~S., \& {Hirano}, S. 2017, \bibinfo{title}{{Formation of intermediate-mass black holes through runaway collisions in the first star clusters},} \mnras, 472, 1677, \dodoi{10.1093/mnras/stx2044}

\bibitem[{M. {Sasaki} {et~al.}(2018){Sasaki}, {Suyama}, {Tanaka}, \& {Yokoyama}}]{SasakiPBH2018CQGra..35f3001S}
{Sasaki}, M., {Suyama}, T., {Tanaka}, T., \& {Yokoyama}, S. 2018, \bibinfo{title}{{Primordial black holes{\textemdash}perspectives in gravitational wave astronomy},} Classical and Quantum Gravity, 35, 063001, \dodoi{10.1088/1361-6382/aaa7b4}

\bibitem[{A.~T.~P. {Schauer} {et~al.}(2023){Schauer}, {Boylan-Kolchin}, {Colston}, {Sameie}, {Bromm}, {Bullock}, \& {Wetzel}}]{Schauer2023}
{Schauer}, A. T.~P., {Boylan-Kolchin}, M., {Colston}, K., {et~al.} 2023, \bibinfo{title}{{Dwarf Galaxy Formation with and without Dark Matter-Baryon Streaming Velocities},} \apj, 950, 20, \dodoi{10.3847/1538-4357/accc2c}

\bibitem[{A.~T.~P. {Schauer} {et~al.}(2019){Schauer}, {Glover}, {Klessen}, \& {Ceverino}}]{Schauer2019MNRAS.484.3510S}
{Schauer}, A. T.~P., {Glover}, S. C.~O., {Klessen}, R.~S., \& {Ceverino}, D. 2019, \bibinfo{title}{{The influence of streaming velocities on the formation of the first stars},} \mnras, 484, 3510, \dodoi{10.1093/mnras/stz013}

\bibitem[{A.~T.~P. {Schauer} {et~al.}(2017){Schauer}, {Regan}, {Glover}, \& {Klessen}}]{Schauer2017MNRAS.471.4878S}
{Schauer}, A. T.~P., {Regan}, J., {Glover}, S. C.~O., \& {Klessen}, R.~S. 2017, \bibinfo{title}{{The formation of direct collapse black holes under the influence of streaming velocities},} \mnras, 471, 4878, \dodoi{10.1093/mnras/stx1915}

\bibitem[{A. {Sesana}(2013){Sesana}}]{Sesana2013MNRAS}
{Sesana}, A. 2013, \bibinfo{title}{{Systematic investigation of the expected gravitational wave signal from supermassive black hole binaries in the pulsar timing band.},} \mnras, 433, L1, \dodoi{10.1093/mnrasl/slt034}

\bibitem[{M. {Shibata} {et~al.}(2025){Shibata}, {Fujibayashi}, {Jockel}, \& {Kawaguchi}}]{Shibata2025}
{Shibata}, M., {Fujibayashi}, S., {Jockel}, C., \& {Kawaguchi}, K. 2025, \bibinfo{title}{{Threshold Mass of the General-relativistic Instability for Supermassive Star Cores},} \apj, 978, 58, \dodoi{10.3847/1538-4357/ad93a4}

\bibitem[{M. {Shibata} \& S.~L. {Shapiro}(2002){Shibata} \& {Shapiro}}]{Shibata2002_DCBH}
{Shibata}, M., \& {Shapiro}, S.~L. 2002, \bibinfo{title}{{Collapse of a Rotating Supermassive Star to a Supermassive Black Hole: Fully Relativistic Simulations},} \apjl, 572, L39, \dodoi{10.1086/341516}

\bibitem[{J. {Silk}(2013){Silk}}]{Silk2013ApJ...772..112S}
{Silk}, J. 2013, \bibinfo{title}{{Unleashing Positive Feedback: Linking the Rates of Star Formation, Supermassive Black Hole Accretion, and Outflows in Distant Galaxies},} \apj, 772, 112, \dodoi{10.1088/0004-637X/772/2/112}

\bibitem[{P.~A. {Solar} {et~al.}(2025){Solar}, {Reinoso}, {Schleicher}, {Klessen}, \& {Banerjee}}]{Solar2025}
{Solar}, P.~A., {Reinoso}, B., {Schleicher}, D.~R.~G., {Klessen}, R.~S., \& {Banerjee}, R. 2025, \bibinfo{title}{{Formation of supermassive stars in the first stellar clusters: Dependence on the gas temperature},} \aap, 699, A64, \dodoi{10.1051/0004-6361/202450903}

\bibitem[{V. {Springel}(2005){Springel}}]{springel2005cosmological}
{Springel}, V. 2005, \bibinfo{title}{{The cosmological simulation code GADGET-2},} \mnras, 364, 1105, \dodoi{10.1111/j.1365-2966.2005.09655.x}

\bibitem[{A. {Stacy} {et~al.}(2011){Stacy}, {Bromm}, \& {Loeb}}]{Stacy2011}
{Stacy}, A., {Bromm}, V., \& {Loeb}, A. 2011, \bibinfo{title}{{Effect of Streaming Motion of Baryons Relative to Dark Matter on the Formation of the First Stars},} \apjl, 730, L1, \dodoi{10.1088/2041-8205/730/1/L1}

\bibitem[{S.~W. {Stahler} {et~al.}(1980){Stahler}, {Shu}, \& {Taam}}]{Stahler1980}
{Stahler}, S.~W., {Shu}, F.~H., \& {Taam}, R.~E. 1980, \bibinfo{title}{{The evolution of protostars. I - Global formulation and results},} \apj, 241, 637, \dodoi{10.1086/158377}

\bibitem[{M. {Suazo} {et~al.}(2019){Suazo}, {Prieto}, {Escala}, \& {Schleicher}}]{Suazo2019SMBH}
{Suazo}, M., {Prieto}, J., {Escala}, A., \& {Schleicher}, D. R.~G. 2019, \bibinfo{title}{{The Role of Gas Fragmentation During the Formation of Supermassive Black Holes},} \apj, 885, 127, \dodoi{10.3847/1538-4357/ab45eb}

\bibitem[{K. {Sugimura} {et~al.}(2014){Sugimura}, {Omukai}, \& {Inoue}}]{Sugimura2014MNRAS.445..544S}
{Sugimura}, K., {Omukai}, K., \& {Inoue}, A.~K. 2014, \bibinfo{title}{{The critical radiation intensity for direct collapse black hole formation: dependence on the radiation spectral shape},} \mnras, 445, 544, \dodoi{10.1093/mnras/stu1778}

\bibitem[{V. {Takhistov} {et~al.}(2022){Takhistov}, {Lu}, {Gelmini}, {Hayashi}, {Inoue}, \& {Kusenko}}]{Takhistov2022}
{Takhistov}, V., {Lu}, P., {Gelmini}, G.~B., {et~al.} 2022, \bibinfo{title}{{Interstellar gas heating by primordial black holes},} \jcap, 2022, 017, \dodoi{10.1088/1475-7516/2022/03/017}

\bibitem[{A.~J. {Taylor} {et~al.}(2025){Taylor}, {Finkelstein}, {Kocevski}, {Jeon}, {Bromm}, {Amor{\'\i}n}, {Arrabal Haro}, {Backhaus}, {Bagley}, {Banados}, {Bhatawdekar}, {Brooks}, {Calabr{\`o}}, {Ch{\'a}vez Ortiz}, {Cheng}, {Cleri}, {Cole}, {Davis}, {Dickinson}, {Donnan}, {Dunlop}, {Ellis}, {Fern{\'a}ndez}, {Fontana}, {Fujimoto}, {Giavalisco}, {Grazian}, {Guo}, {Hathi}, {Holwerda}, {Hirschmann}, {Inayoshi}, {Kartaltepe}, {Khusanova}, {Koekemoer}, {Kokorev}, {Larson}, {Leung}, {Lucas}, {McLeod}, {Napolitano}, {Onoue}, {Pacucci}, {Papovich}, {P{\'e}rez-Gonz{\'a}lez}, {Pirzkal}, {Somerville}, {Trump}, {Wilkins}, {Yung}, \& {Zhang}}]{Taylor2025ApJ...986..165T}
{Taylor}, A.~J., {Finkelstein}, S.~L., {Kocevski}, D.~D., {et~al.} 2025, \bibinfo{title}{{Broad-line AGNs at 3.5 < z < 6: The Black Hole Mass Function and a Connection with Little Red Dots},} \apj, 986, 165, \dodoi{10.3847/1538-4357/add15b}

\bibitem[{D. {Toyouchi} {et~al.}(2021){Toyouchi}, {Inayoshi}, {Hosokawa}, \& {Kuiper}}]{Toyouchi2021IMBH}
{Toyouchi}, D., {Inayoshi}, K., {Hosokawa}, T., \& {Kuiper}, R. 2021, \bibinfo{title}{{Super-Eddington Mass Growth of Intermediate-mass Black Holes Embedded in Dusty Circumnuclear Disks},} \apj, 907, 74, \dodoi{10.3847/1538-4357/abcfc2}

\bibitem[{D. {Toyouchi} {et~al.}(2023){Toyouchi}, {Inayoshi}, {Li}, {Haiman}, \& {Kuiper}}]{Toyouchi2023}
{Toyouchi}, D., {Inayoshi}, K., {Li}, W., {Haiman}, Z., \& {Kuiper}, R. 2023, \bibinfo{title}{{Radiative feedback on supermassive star formation: the massive end of the Population III initial mass function},} \mnras, 518, 1601, \dodoi{10.1093/mnras/stac3191}

\bibitem[{M. Tremmel {et~al.}(2017)Tremmel, Karcher, Governato, Volonteri, Quinn, Pontzen, Anderson, \& Bellovary}]{tremmel2017romulus}
Tremmel, M., Karcher, M., Governato, F., {et~al.} 2017, \bibinfo{title}{The Romulus cosmological simulations: a physical approach to the formation, dynamics and accretion models of SMBHs,} \mnras, 470, 1121

\bibitem[{A. {Trinca} {et~al.}(2022){Trinca}, {Schneider}, {Valiante}, {Graziani}, {Zappacosta}, \& {Shankar}}]{Trinca2022MNRAS.511..616T}
{Trinca}, A., {Schneider}, R., {Valiante}, R., {et~al.} 2022, \bibinfo{title}{{The low-end of the black hole mass function at cosmic dawn},} \mnras, 511, 616, \dodoi{10.1093/mnras/stac062}

\bibitem[{H. {{\"U}bler} {et~al.}(2024){{\"U}bler}, {Maiolino}, {P{\'e}rez-Gonz{\'a}lez}, {D'Eugenio}, {Perna}, {Curti}, {Arribas}, {Bunker}, {Carniani}, {Charlot}, {Rodr{\'\i}guez Del Pino}, {Baker}, {B{\"o}ker}, {Cresci}, {Dunlop}, {Grogin}, {Jones}, {Kumari}, {Lamperti}, {Laporte}, {Marshall}, {Mazzolari}, {Parlanti}, {Rawle}, {Scholtz}, {Venturi}, \& {Witstok}}]{UblerAGN2024MNRAS.531..355U}
{{\"U}bler}, H., {Maiolino}, R., {P{\'e}rez-Gonz{\'a}lez}, P.~G., {et~al.} 2024, \bibinfo{title}{{GA-NIFS: JWST discovers an offset AGN 740 million years after the big bang},} \mnras, 531, 355, \dodoi{10.1093/mnras/stae943}

\bibitem[{L. {Valtaoja} {et~al.}(1989){Valtaoja}, {Valtonen}, \& {Byrd}}]{ValtaojaSMBHB1989ApJ...343...47V}
{Valtaoja}, L., {Valtonen}, M.~J., \& {Byrd}, G.~G. 1989, \bibinfo{title}{{Binary Pairs of Supermassive Black Holes: Formation in Merging Galaxies},} \apj, 343, 47, \dodoi{10.1086/167683}

\bibitem[{L. {Wang} {et~al.}(2022){Wang}, {Tanikawa}, \& {Fujii}}]{Wang2022}
{Wang}, L., {Tanikawa}, A., \& {Fujii}, M. 2022, \bibinfo{title}{{Gravitational wave of intermediate-mass black holes in Population III star clusters},} \mnras, 515, 5106, \dodoi{10.1093/mnras/stac2043}

\bibitem[{Z. {Wang} {et~al.}(2025){Wang}, {Ma}, {Li}, {Cai}, {Wang}, \& {Wu}}]{Wang2025TDE250418144W}
{Wang}, Z., {Ma}, Y., {Li}, Y., {et~al.} 2025, \bibinfo{title}{{The Role of Population III Star Tidal Disruption Events in Black Hole Growth at the Cosmic Dawn},} arXiv e-prints, arXiv:2504.18144, \dodoi{10.48550/arXiv.2504.18144}

\bibitem[{J.~R. {Westernacher-Schneider} {et~al.}(2022){Westernacher-Schneider}, {Zrake}, {MacFadyen}, \& {Haiman}}]{Westernacher-Schneider2022PhRvD.106j3010W}
{Westernacher-Schneider}, J.~R., {Zrake}, J., {MacFadyen}, A., \& {Haiman}, Z. 2022, \bibinfo{title}{{Multiband light curves from eccentric accreting supermassive black hole binaries},} \prd, 106, 103010, \dodoi{10.1103/PhysRevD.106.103010}

\bibitem[{S.~D.~M. {White} \& C.~S. {Frenk}(1991){White} \& {Frenk}}]{White1991ApJ...379...52W}
{White}, S. D.~M., \& {Frenk}, C.~S. 1991, \bibinfo{title}{{Galaxy Formation through Hierarchical Clustering},} \apj, 379, 52, \dodoi{10.1086/170483}

\bibitem[{S.~D.~M. {White} \& M.~J. {Rees}(1978){White} \& {Rees}}]{White1978MNRAS.183..341W}
{White}, S.~D.~M., \& {Rees}, M.~J. 1978, \bibinfo{title}{{Core condensation in heavy halos: a two-stage theory for galaxy formation and clustering.},} \mnras, 183, 341, \dodoi{10.1093/mnras/183.3.341}

\bibitem[{J. {Wolcott-Green} {et~al.}(2011){Wolcott-Green}, {Haiman}, \& {Bryan}}]{Wolcott2011MNRAS.418..838W}
{Wolcott-Green}, J., {Haiman}, Z., \& {Bryan}, G.~L. 2011, \bibinfo{title}{{Photodissociation of H$_{2}$ in protogalaxies: modelling self-shielding in three-dimensional simulations},} \mnras, 418, 838, \dodoi{10.1111/j.1365-2966.2011.19538.x}

\bibitem[{T.~E. {Woods} {et~al.}(2024){Woods}, {Patrick}, {Whalen}, \& {Heger}}]{Woods2021}
{Woods}, T.~E., {Patrick}, S., {Whalen}, D.~J., \& {Heger}, A. 2024, \bibinfo{title}{{On the Formation and Interaction of Multiple Supermassive Stars in Cosmological Flows},} \apj, 960, 59, \dodoi{10.3847/1538-4357/ad054a}

\bibitem[{T.~E. {Woods} {et~al.}(2019){Woods}, {Agarwal}, {Bromm}, {Bunker}, {Chen}, {Chon}, {Ferrara}, {Glover}, {Haemmerl{\'e}}, {Haiman}, {Hartwig}, {Heger}, {Hirano}, {Hosokawa}, {Inayoshi}, {Klessen}, {Kobayashi}, {Koliopanos}, {Latif}, {Li}, {Mayer}, {Mezcua}, {Natarajan}, {Pacucci}, {Rees}, {Regan}, {Sakurai}, {Salvadori}, {Schneider}, {Surace}, {Tanaka}, {Whalen}, \& {Yoshida}}]{Woods2019}
{Woods}, T.~E., {Agarwal}, B., {Bromm}, V., {et~al.} 2019, \bibinfo{title}{{Titans of the early Universe: The Prato statement on the origin of the first supermassive black holes},} \pasa, 36, e027, \dodoi{10.1017/pasa.2019.14}

\bibitem[{H. {Xu} {et~al.}(2023){Xu}, {Chen}, {Guo}, {Jiang}, {Wang}, {Xu}, {Xue}, {Caballero}, {Yuan}, {Xu}, {Wang}, {Hao}, {Luo}, {Lee}, {Han}, {Jiang}, {Shen}, {Wang}, {Wang}, {Xu}, {Wu}, {Manchester}, {Qian}, {Guan}, {Huang}, {Sun}, \& {Zhu}}]{CPTA2023RAA}
{Xu}, H., {Chen}, S., {Guo}, Y., {et~al.} 2023, \bibinfo{title}{{Searching for the Nano-Hertz Stochastic Gravitational Wave Background with the Chinese Pulsar Timing Array Data Release I},} Research in Astronomy and Astrophysics, 23, 075024, \dodoi{10.1088/1674-4527/acdfa5}

\bibitem[{Y.~B. {Zel'dovich}(1970){Zel'dovich}}]{Zeldovich1970A&A.....5...84Z}
{Zel'dovich}, Y.~B. 1970, \bibinfo{title}{{Gravitational instability: An approximate theory for large density perturbations.},} \aap, 5, 84

\bibitem[{Y.~B. {Zel'dovich} \& I.~D. {Novikov}(1967){Zel'dovich} \& {Novikov}}]{Zeldovich1967SvA....10..602Z}
{Zel'dovich}, Y.~B., \& {Novikov}, I.~D. 1967, \bibinfo{title}{{The Hypothesis of Cores Retarded during Expansion and the Hot Cosmological Model},} \sovast, 10, 602

\bibitem[{S. {Zhang} {et~al.}(2024{\natexlab{a}}){Zhang}, {Bromm}, \& {Liu}}]{Zhang:2024PBH}
{Zhang}, S., {Bromm}, V., \& {Liu}, B. 2024{\natexlab{a}}, \bibinfo{title}{{How Do Primordial Black Holes Change the Halo Mass Function and Structure?},} \apj, 975, 139, \dodoi{10.3847/1538-4357/ad7b0d}

\bibitem[{S. {Zhang} {et~al.}(2024{\natexlab{b}}){Zhang}, {Liu}, \& {Bromm}}]{Zhang2024MNRAS.528..180Z}
{Zhang}, S., {Liu}, B., \& {Bromm}, V. 2024{\natexlab{b}}, \bibinfo{title}{{Distinguishing the impact and signature of black holes from different origins in early cosmic history},} \mnras, 528, 180, \dodoi{10.1093/mnras/stad3986}

\bibitem[{S. Zhang {et~al.}(2025)Zhang, Liu, \& Bromm}]{zhang_2025_17025634}
Zhang, S., Liu, B., \& Bromm, V. 2025, \bibinfo{title}{PHANTOM: Primordial black Holes And Nonlinear perTurbations fOr siMulations,} Zenodo, \dodoi{10.5281/zenodo.17025634}

\bibitem[{S. {Zhang} {et~al.}(2025){Zhang}, {Liu}, {Bromm}, {Jeon}, {Boylan-Kolchin}, \& {K{\"u}hnel}}]{Zhang2025}
{Zhang}, S., {Liu}, B., {Bromm}, V., {et~al.} 2025, \bibinfo{title}{{How do Massive Primordial Black Holes Impact the Formation of the First Stars and Galaxies?},} \apj, 987, 185, \dodoi{10.3847/1538-4357/adddb4}

\bibitem[{F. {Ziparo} {et~al.}(2025){Ziparo}, {Gallerani}, \& {Ferrara}}]{Ziparo2025JCAP...04..040Z}
{Ziparo}, F., {Gallerani}, S., \& {Ferrara}, A. 2025, \bibinfo{title}{{Primordial black holes as supermassive black hole seeds},} \jcap, 2025, 040, \dodoi{10.1088/1475-7516/2025/04/040}

\bibitem[{F. {Ziparo} {et~al.}(2022){Ziparo}, {Gallerani}, {Ferrara}, \& {Vito}}]{Ziparo2022}
{Ziparo}, F., {Gallerani}, S., {Ferrara}, A., \& {Vito}, F. 2022, \bibinfo{title}{{Cosmic radiation backgrounds from primordial black holes},} \mnras, 517, 1086, \dodoi{10.1093/mnras/stac2705}

\end{thebibliography}
\bibliographystyle{aasjournal}



\end{document}